\def\conftype{1} 
\def\publish{1}

\ifcase 
\conftype  
\documentclass[sigconf, 10pt]{acmart}
\usepackage[english]{babel}
\renewcommand\footnotetextcopyrightpermission[1]{} 
\setcopyright{none}
\acmDOI{}
\acmISBN{}
\acmPrice{}
\or 
\documentclass[letterpaper,twocolumn,10pt]{article}
\usepackage{usenix-2020-09}
\fi

\usepackage{titlesec}
\titlespacing*{\subsection}{0pt}{1pt}{1pt}
\titlespacing*{\subsubsection}{1pt}{1pt}{1pt}
\usepackage[skip=5pt]{caption}

\usepackage{stfloats}
\usepackage{natbib}
\usepackage{amssymb}
\usepackage{pifont}
\usepackage{tikz}
\usepackage[linesnumbered,ruled,vlined]{algorithm2e}
\usepackage[noend]{algpseudocode}
\usepackage{array}
\newcolumntype{x}[1]{>{\centering\arraybackslash\hspace{0pt}}p{#1}}
\usepackage{amsmath}
\usepackage[inline]{enumitem}
\usepackage{epsfig}
\usepackage{fancyvrb}
\usepackage{graphicx}
\usepackage{booktabs, tabularx}
\usepackage{makecell}
\usepackage[flushleft]{threeparttable}
\usepackage{latexsym, color}
\usepackage{listings}
\usepackage{tikz}
\usepackage{url}
\usepackage[font=normalsize,labelfont=bf]{caption}
\usepackage{hyperref}
\usepackage{cleveref}
\usepackage{tabularx}
\usepackage{subcaption}
\usepackage{xcolor}
\usepackage{multirow}
\usepackage{multicol}
\usepackage{float}
\usepackage{array}
\usepackage{algpseudocode}
\usepackage{setspace}
\usepackage{microtype}
\usepackage{amsmath,amsthm,amssymb,amsfonts}
\usepackage{minted}
\usepackage{float}
\usepackage{placeins}

\crefformat{figure}{Figure~#2#1#3}

\definecolor{commentgreen}{RGB}{2,112,10}
\definecolor{eminence}{RGB}{108,48,130}
\definecolor{weborange}{RGB}{255,165,0}
\definecolor{frenchplum}{RGB}{129,20,83}

\long\def\comment#1{}

\usepackage{cuted}
\usepackage{float}
\usepackage{xspace}

\newcommand{\system}{\textit{$S^2Sim$}}
\newcommand{\ie}{{\em i.e.}}
\newcommand{\eg}{{\em e.g.}}

\newcommand{\para}[1]{\noindent {\bf #1}}

\newcommand{\qiao}[1]{\textcolor{red}{[qiao: #1]}}
\newcommand{\rulan}[1]{\textcolor{purple}{[rulan: #1]}}

\newcommand{\xing}[1]{\textcolor{red}{[xing: #1]}}

\begin{document}

\pagenumbering{gobble}  

\ifcase 
\conftype
\title[Configuration Diagnosis and Repair]{Diagnosing and Repairing Distributed Routing Configurations Using Selective Symbolic Execution}
\or
\title{Diagnosing and Repairing Distributed Routing Configurations Using Selective Symbolic Simulation}
\fi


\author{
\ifcase\publish
    Paper ID: \# 155, \\
	Number of Pages: 12 pages + References (2 pages) + Appendices (3 pages)
  \or
    Rulan Yang $^{1}$, 
    Gao Han $^{1}$,
    Hanyang Shao $^{1}$,
    Xiaoqiang Zheng $^{1}$,
    Xing Fang $^{1}$, 
    Ziyi Wang $^{1}$,
    \\
    Lizhao You $^{1}$\footnotemark[1], 
    Ruiting Zhou $^{2}$,
    Linghe Kong $^{3}$,
    Ennan Zhai $^{4}$,
    Qiao Xiang $^{1}$\footnotemark[1],   
    Jiwu Shu $^{1,}$$^{5}$
    \\
    $^{1}$Xiamen University, 
    $^{2}$Southeast University,
    $^{3}$Shanghai Jiao Tong University,
    \\
    $^{4}$Alibaba Cloud,
    $^{5}$Minjiang University
  \fi	
}

\ifcase 
\conftype
\ifcase
\publish
\renewcommand{\shortauthors}{Anonymous authors}
\or
\renewcommand{\shortauthors}{Yang et al.}
\fi
\fi

\maketitle
\renewcommand{\thefootnote}{\fnsymbol{footnote}}
\footnotetext[1]{Lizhao You and Qiao Xiang are co-corresponding authors.}
\renewcommand{\thefootnote}{\arabic{footnote}}  

\newcommand{\todo}[1]{{\color{red}#1}} 

\begin{abstract}


Although substantial progress has been made in automatically verifying whether
distributed routing configurations comply with certain intents,
diagnosing and repairing configuration errors remains manual and
time-consuming. To fill this gap, we propose \system{}, a novel system
for automatic routing configuration diagnosis and repair. 
Our key insight is that by deriving a set of contracts that guarantees an intent-compliant variant of the erroneous configuration, we can
systematically check for all contract violations in the configuration
via symbolic simulation to pinpoint and repair the errors. 
\system{} also introduces a series of extensions to support complex configurations
(\eg, ACL, route aggregation and multi-path routing), networks (\eg,
underlay and overlay networks), and intents (\eg, $k$-link failure tolerance). 
We fully implement \system{}
and evaluate its performance using real configurations from two major
providers and synthesized configurations composed from their real errors
and real-world topologies with different scales $O(10)$ to $O(1000)$. 
Results show that \system{} accurately and efficiently diagnoses and repairs real configuration
errors (\ie, up to 20 seconds in real networks of $O(100)$ nodes and up to 15 minutes in synthesized networks of $O(1000)$ nodes).

\end{abstract}

\vspace{-10pt}
\section{Introduction}
\vspace{-8pt}

Many network configuration verification tools, also called control plane
verification (CPV) tools, have been designed and
deployed~\cite{feamster2005detecting, fayaz2016efficient, fogel2015general,
beckett2017general,gember2016fast,beckett2018control,beckett2019abstract,giannarakis2019efficient,ye2020hoyan,abhashkumar2020tiramisu,zhang2022dna,liu2017crystalnet,lopes2019fast,zhang2022sre,
brown2023batfish, fang2024network, xu2024relational, wang2024expresso,
li2024general} . They analyze the configurations of routers and switches to
determine whether these configurations, when deployed in production networks,
would yield a data plane that satisfied a series of pre-specified intents (\eg,
reachability, blackhole freeness, waypointing, and loop freeness).

\para{Diagnosing and repairing configuration errors remains manual,
time-consuming and error-prone.} Despite the substantial progress in
expanding the capability of CPV and accelerating them, once a set of
configurations are deemed erroneous, it is still up to network operators to
manually locate the errors and repair them. Although some CPV tools (\eg,
Minesweeper~\cite{beckett2017general}) also return a counter example to users,
but repairing the configurations, in the worst case, requires enumerating all
possible counter examples, which is NP-complete.

Some studies have attempted to tackle the problems of diagnosing and repairing
network configurations. Earlier data provenance-based
tools~\cite{zhou2010efficient, chen2018data, loo2009declarative,
wu2014diagnosing, wu2017automated} analyze the causal relationship among network events to find the root causes (\eg, link failures) of network behaviors
(\eg, updates of forwarding rules). However, they require re-implementing
routing protocols and their configurations in a declarative datalog dialect
(\eg, NDlog~\cite{zhou2010efficient}), which brings additional learning curves for operators
and limits their deployment. They also could not identify latent errors (\ie,
errors that have not caused a failure yet, but will do once other configurations are
updated.)

\para{Recent tools based on program analysis cannot accurately localize and
repair configuration errors~(\S\ref{sec:limitations}).} 
Recent tools~\cite{gember2017automatically, gember2022localizing, acr-hotnets24} 
resort to program analysis to diagnose or repair configuration errors, but have
different limitations. 
CEL~\cite{gember2022localizing} extends Minesweeper to model network
configurations and intents into an SMT formula, and computes its minimal
correction set (\ie, a subset of constraints whose removal would allow the
remaining of the formula to become satisfied) as the configuration errors. However, it cannot support AS-path related configurations (\eg,
a route filter matching on ASes in the AS path) due to the path encoding
explosion of the SMT formula. CPR~\cite{gember2017automatically} models network
configurations using an abstract representation and repairs them using constraint
programming, but it cannot support local preference-related
configurations due to its graph-based abstraction.
ACR~\cite{acr-hotnets24} applies a spectrum-based method to rank potential erroneous
lines and uses an experience-based method to find a possible repair. Not only
could it require multiple trial-and-error without success, it also relies on
test coverage based tools (\eg, NetCov~\cite{xu2023test}) to first identify
some lines as candidates, which could miss real errors even in a simple
network.

In this paper, we systematically investigate the important problem of how to
accurately and efficiently diagnose and repair errors in network configurations. 



\para{Proposal: Explore variants of the original erroneous configuration to find
an intent-compliant one, and compare their differences for error diagnosis and
repair.}
Instead of analyzing the given configuration to find out what went wrong,
we resort to a different mindset: if we can find an intent-compliant
configuration that is similar to the original erroneous one, their differences
essentially suggest the error and the repair. At first glance, this 
appears to be a \textit{catch-22}: in order to design a tool to
diagnose and repair configurations, we must first have a tool that can find an
intent-compliant variant of the original configuration (\ie, updating small
portion of the original one to make it intent-compliant).

\para{Key insight: Find a set of contracts that guarantee an intent-compliant
variant, and detect where they were breached in the original configuration as
errors via symbolic simulation~(\S\ref{sec:overview}).}
We circumvent the paradox above by not searching for a concrete configuration,
but a set of \textit{contracts} such that if a configuration satisfy them all, it
must be intent-compliant. Informally, a contract is a Boolean predicate that
specifies the behavior of a router (\ie, whether to peer with another router and
whether to discard/prefer/announce a route). By simulating the erroneous
configuration symbolically to identify all the snippets that could violate the
intent-compliant contracts, we can localize and repair the errors in the
configuration. While a previous position paper by Yang et al. \cite{yang2023diagnosing} also
suggests the promise of diagnosing and repairing network
configuration using symbolic simulation, it fell short in answering several important
questions, including (1) how to support complex configurations (\eg, access control lists, route aggregation, and multi-path routing), 
(2) how to cope with complex networks where multiple routing protocols (\eg, path-vector and link-state) co-exist,
and (3) how to diagnose and repair the fault-tolerance of configuration.
To this end, we design \system{}, a generic, efficient, lightweight tool for
network configuration diagnosis and repair with three key designs (D1-D3).

\para{D1: Derive intent-compliant contracts from erroneous data plane and
intents, and check their violations with selective symbolic simulation~(\S\ref{sec:basic}).}
After finding a configuration is erroneous via simulation, \system{} designs an
algorithm based on DFA-multiplication to compute an intent-compliant data plane
with a small difference from the erroneous one. It derives a set of contracts
that is sufficient and necessary to yield this data plane.
Next, \system{} starts another simulation of the erroneous configuration, which
is selective and symbolic. At each step, if a contract is violated,
\system{} forces the simulation to obey this contract and switch the simulation to a symbolic configuration variant.
Because the simulation obeys all the derived contracts, it eventually yields the intent-compliant data plane. As such, \system{}
can localize the errors in the original configuration by mapping the violated contracts to erroneous configuration snippets and repair them using an algorithm based on constraint
programming. \system{} also extends to support a wide range of 
complex configurations such as route aggregation, access control lists (ACL), and multi-path routing.



\para{D2: Use an assume-guarantee approach to diagnosis and repair a network with multiple
routing protocols~(\S\ref{sec:multi-protocol}).} 
For networks running different routing protocols in underlay and overlay (\eg,
OSPF as underlay and BGP as overlay), \system{} diagnosis and repair layers
separately for modularity. It first assumes the proper functioning of the
underlay network (\eg, reachability between overlay-neighboring routers) and
diagnoses and repairs the overlay network to be intent-compliant. It then uses
the assumption as the intents for the underlay network and repeats the diagnosis
and repair process. 

\para{D3: Derive fault-tolerant contracts for configuration
fault-tolerance~(\S\ref{sec:k-failure}).}
To diagnose and repair configuration so that it is intent-compliant even when
there are up to $k$ link failures in the network, \system{} extends the
basic design to compute an intent-compliant data plane that is also fault-tolerant, and derives fault-tolerant contracts accordingly. It then incorporates the selective symbolic simulation to check for contract violations under
fault-tolerant contracts and use these violations for error
localization and repair.

\para{Evaluation with production data~(\S\ref{sec:eval}).} 
We implement a closed-source version of \system{} for a major provider as a
plug-in of its in-house simulation CPV tool, and an open-source prototype\footnote{https://github.com/sngroup-xmu/S2Sim} 
as a plugin of Batfish~\cite{fogel2015general}. We evaluate \system{} with (1)
functionality demos on a small network~\cite{demo-2024}; (2) experiments using
real configurations from production networks of two major providers; and (3) experiments using configurations synthesized from the providers'
real errors and real-world topologies of different scales $O(10)$ to $O(1000)$.
Results show that \system{} accurately and efficiently diagnoses and repairs configuration errors in real networks (\ie, up to 20 seconds in DC-WAN and IPRANs with $O(100)$ nodes) and synthesized large-scale networks (\ie, up to 15 minutes in IPRANs and DCNs with $O(1000)$ nodes).

\vspace{-10pt}
\section{Why Existing Tools Do Not Work}\label{sec:limitations}
\vspace{-10pt}

In this section, we demonstrate the limitations of related tools through an
experiment in a simple network. The screenshots of outputs of all our experiments
are in Appendix~\ref{appendix}.

\begin{figure}
    \centering
    \includegraphics[width=\linewidth]{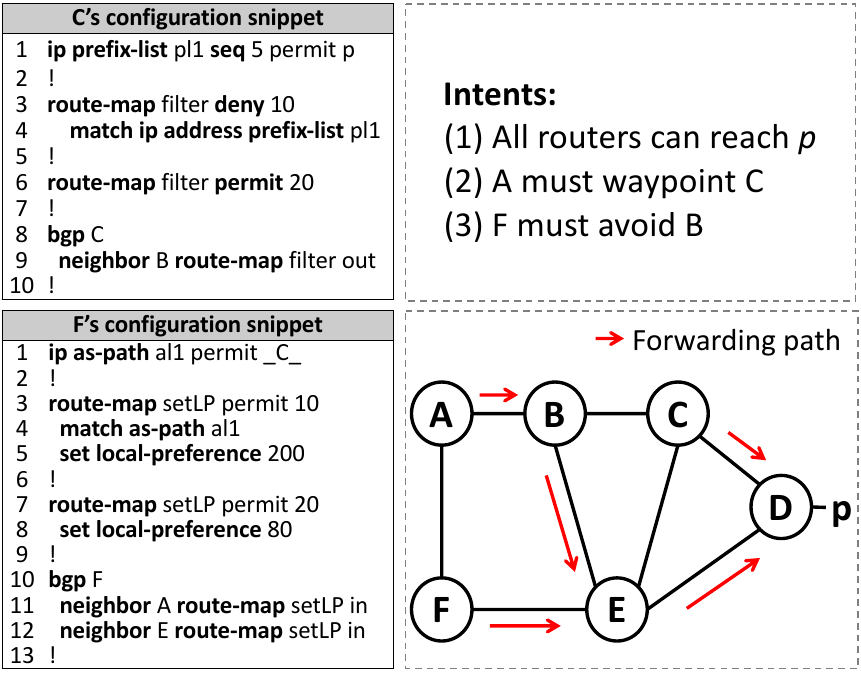}
    \caption{The example network.}
    \label{fig:example}
    \vspace{-20pt}
\end{figure}

\para{An example network.} 
Consider a small network of six routers running BGP in Figure~\ref{fig:example}.
For simplicity, we use the ID of routers as their AS numbers. The destination IP
prefix $p$ is at router $D$.  The intents of the operator are: (1) all routers can reach prefix $p$; (2) router $A$ must waypoint router $C$; and (3) router $F$ must avoid router $B$. 
All the routers use the default BGP configuration except for the snippets of router $C$ and $F$ shown in the figure, where $C$ does not send route with prefix $p$ to $B$, and $F$ sets a higher local preference for all routes whose AS-path contains $C$. Because it is such a small network, we can simulate its execution on paper and observe that while all the routers can reach $p$, the path of $A$ does not satisfy intent 2.

\para{Errors in the configuration: the ground truth.}
There are two errors in the configuration. 
One may quickly observe that one is the export policy at $C$ that denies routes with prefix $p$ to $B$. However, removing it only partially fixes the
configuration. For example, if $C$ removes this policy, $B$ will receive and prefer $[B, C, D]$ over $[B, E, D]$ as its path to reach $p$ because $C$ has a lower ID
than $E$, and $A$ will also use path $[A, B, C, D]$ to reach $p$.
But $F$ will also switch to $[F, A, B, C, D]$ because it has a policy to prefer the path that contains AS $C$ over all other paths. Therefore, router $F$'s
policy to prefer the AS-path that contains $C$ is also an error and needs to be removed.

\para{Outputs of representative configuration verification/diagnosis/repair tools.}
We select five recent tools to examine the configuration of this small network against the three intents. To make sure the experiment is fair, we use the open-source prototypes from the authors. We find that although some of them can determine that the configuration is incorrect, none of them can correctly locate and repair the two errors.

\para{Batfish~\cite{fogel2015general, brown2023batfish} correctly determines the configuration is erroneous but cannot locate the errors.} As a simulation-based CPV, Batfish computes the paths of each router based on their configurations and alerts us that these paths violate intent 2  (Figure~\ref{fig:Batfish-output} in Appendix~\ref{appendix}). However, it neither identifies the incorrect configuration nor suggests a fix.

\para{Minesweeper~\cite{beckett2017general} also correctly determines the configuration is erroneous and cannot locate the errors.} As an SMT-based CPV, Minesweeper alerts us that the configuration is erroneous and provides a corresponding counter
example (Figure~\ref{fig:MS-output} in Appendix~\ref{appendix}).  
Likewise, it provides no suggestions for localizing or repairing the configuration error.

\para{CEL~\cite{gember2022localizing} fails to find all errors in the configuration.} CEL analyzes the SMT formula constructed by Minesweeper and computes its minimal correction set (MCS) as the errors. However, CEL only finds the errors at router C in this example. This is because CEL's encoding is based on Minesweeper, which does not support AS-path related configuration because of the path explosion. (Figure~\ref{fig:CEL-output} in Appendix~\ref{appendix}). 

\para{CPR~\cite{gember2017automatically} fails to repair any error in this configuration.} CPR uses an abstract graph representation to model the network configuration and finds the repair using constraint programming. CPR identified the configuration as erroneous, but incorrectly repaired the error by adding an ACL in router $A$ to block all traffic from $A$ to $D$ (Figure~\ref{fig:CPR-output} in Appendix~\ref{appendix}). 
This is because CPR's graph model incorrectly determines that $A$ can reach $C$ and cannot support the local preference-related configurations.


\para{ACR~\cite{acr-hotnets24} cannot locate or repair any error.} ACR adopts a trial-and-error approach. In each
iteration, it applies a spectrum-based method to rank potential
erroneous lines, tries to repair them using an experience-based method, and
validates the repair using CPV (\eg, Batfish). Although it
does not have an open-source prototype, it suggests that it uses a configuration
test coverage tool (\ie, NetCov~\cite{xu2023test}) to first identify configuration lines related to the errors as candidates. Therefore, we implement ACR by ourselves with NetCov as a key component. In this example,
however, when we send NetCov a test case of path from $A$ to $D$, the candidate
lines NetCov returns do not include any route policies at $C$ (Figure~\ref{fig:netcov-output} in Appendix~\ref{appendix}), the
actual errors in the network. This is because NetCov leverages positive provenance to identify the related configuration for existing routes, but it cannot locate the configuration contributing to the nonexistence of a route. As a result, ACR cannot locate or repair the configuration after repeating the trial-and-error process multiple times.

\para{Takeaway: an accurate tool for configuration diagnosis and repair is still
missing.} 
Representative existing tools have different limitations,
making it difficult for them to locate or repair the configuration errors, even in a simple network with simple configurations.

\vspace{-10pt}
\section{\system{} in a Nutshell}\label{sec:overview}
\vspace{-5pt}

We use the same
example in Figure~\ref{fig:example} to illustrate the key insight and basic
workflow of how \system{} accurately diagnoses and repairs network
configuration.
We also provide an interactive demo~\cite{demo-2024} of this example to demonstrate the functionality of our tool.

\begin{figure}
    \centering
    \includegraphics[width=\linewidth]{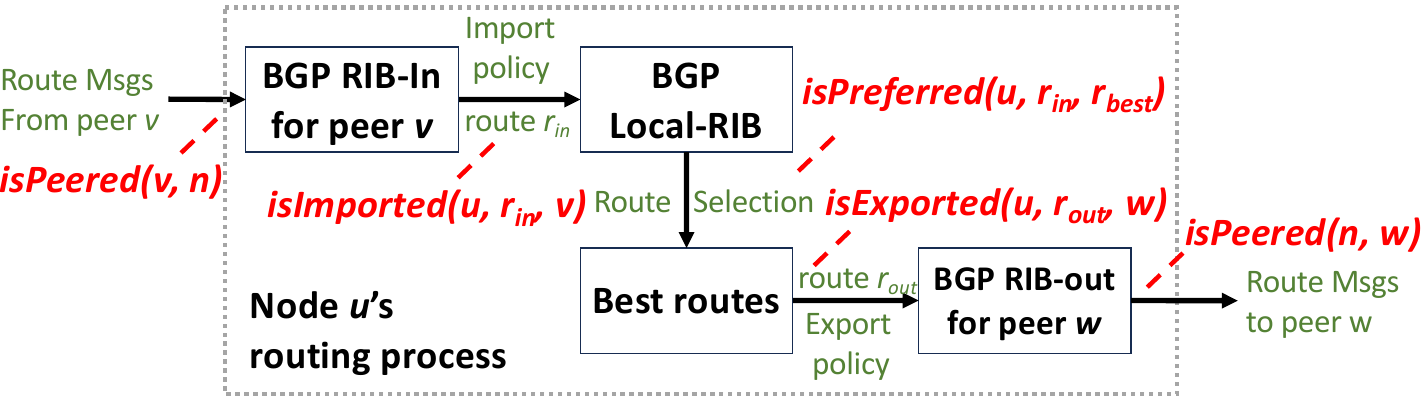}
    \caption{The BGP process of a router. The configuration determines the routing contracts (shown in red).}
    \label{fig:BGP-program}
    \vspace{-20pt}
\end{figure}

\subsection{Contracts}\label{sec:contract}
A contract is a predicate that specifies the behavior of a router. 
To be concrete, a routing process runs as an event-driven program. It receives route updates (\eg, BGP and link-state route announcement) and processes them through a series of conditional statements. Each Boolean condition in this program is a \textit{contract}. 
For example, Figure~\ref{fig:BGP-program} provides a basic overview of a BGP process, comprising four types of contracts: $isPeered$, $isImported$, $isPreferred$, and $isExported$, each governing specific routing behavior within the process. Table~\ref{tab: contracts} provides a summary of common contracts.

\para{Key observation: network configuration decides the values of contracts.}
Given a network configuration script, it does not change the structure of the
event-driven program, but only the value of its contracts. Taking the route policy of router $C$ in Figure~\ref{fig:example} as an example, this contract is to set the $isExported$ value to be $false$ for all routes with prefix $p$ to router $B$.

\vspace{1em}
\subsection{Workflow}\label{sec:workflow}
\vspace{1em}

The basic workflow of \system{} consists of three steps: (1) find a set of contracts that ensure an intent-compliant configuration; (2) simulate the original configuration symbolically and selectively to find all violations of these contracts; (3) locate errors by mapping violated contracts to configuration snippets and compute a conflict-free repair using constraint programming. We illustrate this procedure using the example in Figure~\ref{fig:example}.

\begin{figure}
    \centering
    \includegraphics[width=\linewidth]{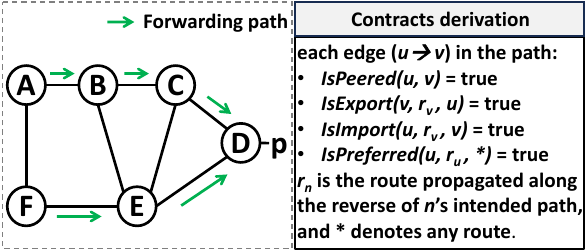}
    \caption{The intent-compliant data plane and contracts of the example network.}
    \label{fig:intent-dp-contract-example}
    \vspace{-10pt}
\end{figure}

\para{Step 1: Derive intent-compliant contracts from the intents and erroneous data plane.} 
Motivated by recent network synthesis (\eg, Propane~\cite{beckett2016don} and
Contra~\cite{hsu2020contra}) and verification tools (\eg, RCDC~\cite{jayaraman2019validating} and Tulkun~\cite{xiang2023beyond}),
we derive intent-compliant contracts from an intent-compliant
data plane. A strawman to find such a data plane is to express intents in a regular expression
and multiply it with the topology (\ie, take their cross product). Although it works, the resulting data plane
could be very different from the one yielded by the original erroneous
configuration. 
For example, paths $[A, F, E, C, D]$ and $[A, B, C, D]$ are both compliant with $A's$ intent.
However, if we derive contracts from the former one and use them to
guide the diagnosis and repair, it could result in a large number of changes of
lines in the configuration (including changing $A$ and $E$'s configurations).

With this in mind, we reuse the intent-compliant part of the erroneous data
plane from the original configuration as a starting point, compute an
intent-compliant data plane with a slight difference from the erroneous one, and
use it to derive intent-compliant contracts. Our intuition is that the smaller the
change in the data plane, the fewer contract violations in the configuration, and
the fewer lines in the configuration need to be updated.
Specifically, we begin with the set of satisfied paths in the original erroneous data plane as our initial path constraints. For each unsatisfied intent, we compute the shortest valid path that satisfies both its regular expression and the current path constraints, and add it to the path constraints. 
During this iteration, if no valid path exists for an intent, we backtrack by gradually relaxing path constraints, \ie, removing one path at a time from path constraints and marking the corresponding intents as unsatisfied, until a valid path for the current intent is found, then proceed to the next intent.
To address the combinatorial complexity introduced by the ordering of path finding and backtracking operations, we introduce two principles in §\ref{sec:derive-contracts}, which ensure that we can efficiently construct an intent-compliant data plane while reusing as many segments of the original data plane as possible.

In the example in Figure~\ref{fig:example}, router $B$, $C$, $E$ and $F$ each has an
intent-compliant path, but $A$ does not. We try to find a valid path for $A$'s intent
(\ie, $A .^* C .^* D$ and loop-free) and uses the paths of
other routers as guidance constraints. 
However, no valid path exists for $A$'s intent. 
Therefore, we remove a satisfied path $[B, E, D]$ from the path constraints and mark $B$'s intent as unsatisfied, this relaxation allows us to find a valid path for $A$ ($[A, B, C, D]$). 
In the next iteration, we reprocess $B$'s intent and find $[B,C,D]$ is already a valid path.
In the end, we get an intent-compliant data plane
in Figure~\ref{fig:intent-dp-contract-example} with only one link different from the erroneous data plane.

\begin{figure}
    \centering
    \includegraphics[width=\linewidth]{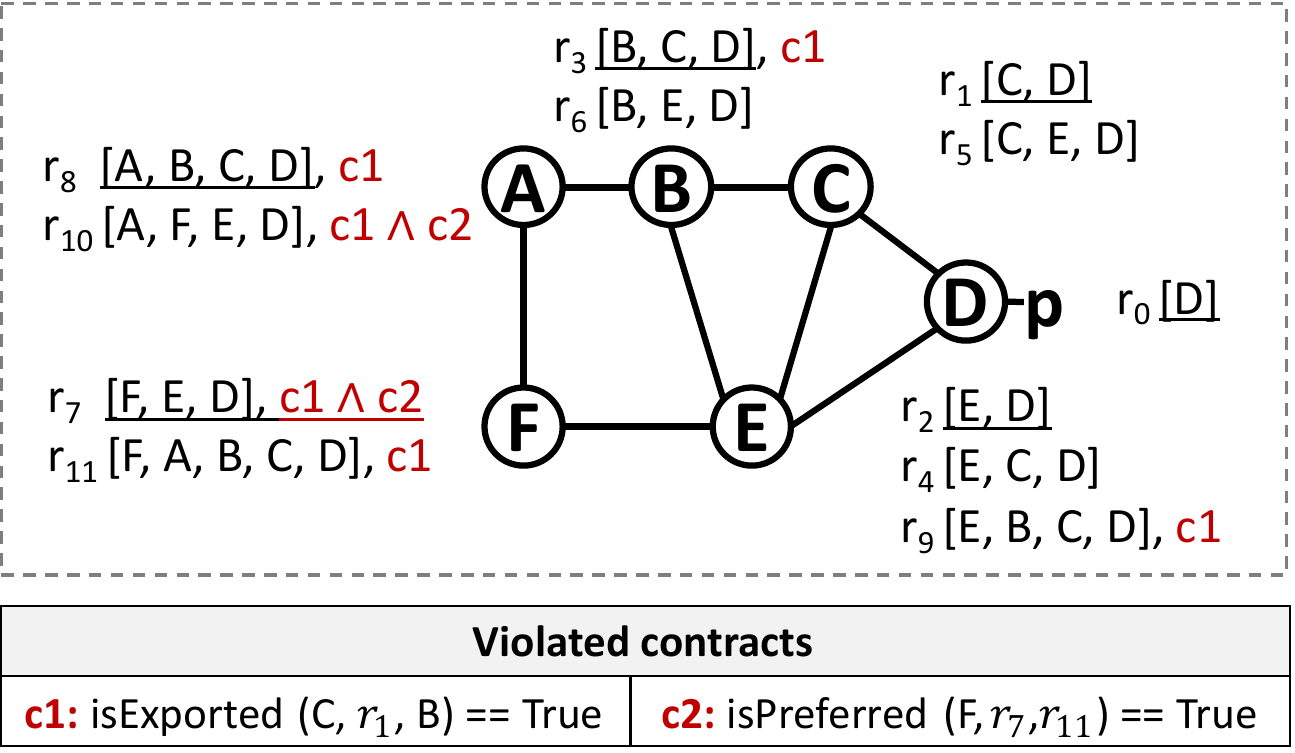}
    \caption{The symbolic simulation of the example network.}
    \label{fig:workflow-c}
    \vspace{-20pt}
\end{figure}

\begin{table*}[t]
 \centering
 \small
\begin{tabularx}{\linewidth}{|
  >{\hsize=0.19\hsize}X |
  >{\hsize=0.4\hsize}X |
  >{\hsize=0.44\hsize}X |}
\hline
\multicolumn{1}{|c|}{\textbf{Contracts}}   & \multicolumn{1}{c|}{\textbf{Descriptions}}                      & \multicolumn{1}{c|}{\textbf{Mapped Configuration Snippets}} \\
\hline

isPeered ($u$, $v$) & 
\makecell{Whether to establish a peer session \\between nodes $u$ and $v$ (for path-vector protocols)} & 
\makecell{Peer snippets on $u$ and $v$}  \\
\hline

isEnabled ($u$, $v$) & 
\makecell{Whether the interfaces between nodes $u$ and $v$ \\ are in the same area (for link-state protocols)} & 
\makecell{Interface snippets on $u$ and $v$} \\
\hline

\multirow{2}{=}{isPreferred ($u$, $r$, $r'$)} & 
\multirow{2}{=}{\makecell{Whether to prefer route $r$ over $r'$ at node $u$}} & 
\makecell{Import policy snippets that match $r$ and $r'$ on $u$} \\
\cline{3-3}
 & & \makecell{Link cost snippets on nodes along paths of $r$ and $r'$} \\
\hline

isExported ($u$, $r$, $v$) & 
\makecell{Whether to export a route $r$ to $v$ at node $u$} & 
\makecell{Export policy snippets on $u$ matching route $r$ for $v$} \\
\hline

isImported ($u$, $r$, $v$) & 
\makecell{Whether to import a route $r$ from $v$ at node $u$} & 
\makecell{Import policy snippets on $u$ matching route $r$ from $v$} \\
\hline

isEqPreferred ($u$, $r$, $r'$) & 
\makecell{Whether to equally prefer routes $r$ and $r'$ at node $u$} & 
\makecell{Import policy snippets matching $r$ and $r'$ and multi-\\path policy on $u$} \\
\hline

isForwardedIn ($u$, $p$, $v$) & 
\makecell{Whether to forward packet $p$ from $v$ at node $u$} & 
\makecell{Inbound ACL snippets on $u$ that match packet $p$ from $v$} \\
\hline

isForwardedOut ($u$, $p$, $v$) & 
\makecell{Whether to forward packet $p$ to $v$ at node $u$} & 
\makecell{Outbound ACL snippets on $u$ that match packet $p$ to $v$} \\
\hline

\end{tabularx}
    \caption{Common contracts and their mapped configuration snippets. }
    \label{tab: contracts}
    \vspace{-20pt}
\end{table*}

\para{From an intent-compliant data plane to intent-compliant contracts.}
The reverse of the data plane computed above (\ie, changing the direction of all
links) provides a set of \textit{sufficient and necessary} conditions for the
routers to yield it. We call them the \textit{intent-compliant contracts}.  To
be concrete, a forwarding path $ [\ldots, R_{i-1}, R_{i}, R_{i+1}, \ldots]$ will
exist in the data plane if and only if for each router $R_i$, it peers with
$R_{i+1}$, imports the route $[R_{i+1}, \ldots]$, prefers it over other routes (\ie, the best route),
and exports it to $R_{i-1}$.
In other words, when $R_i$ processes the route
$[R_{i+1}, \ldots]$, all its contracts (\ie, $isPeered$, $isImported$,
$isPreferred$, and $isExported$) must be $true$. We call such routes intent-compliant routes. Figure~\ref{fig:intent-dp-contract-example} lists the
intent-compliant contracts of this example.

\para{Step 2: Simulate the configuration to identify all violations of
the intent-compliant contracts.}
We start an iterative simulation of the original configuration. At each step,
for each router, we examine the values of its contracts. If they are different
from the values of the intent-compliant contracts, we record this violation,
force the router's behavior to obey the contracts, and continue
the simulation. As such, the simulation switches to a symbolic variant of the
original configuration.  In the end, the simulation yields the intent-compliant
data plane in Step 1 as the output because (1) the whole simulation follows the
intent-compliant contracts as invariants, and (2) the latter are the sufficient
and necessary conditions for generating the intent-compliant data plane. 


Figure~\ref{fig:workflow-c} shows the simulation of our
example. At one step, when $C$ sends $[C, D]$ to $B$, the original configuration
decides that $C$ will set the corresponding $isExported$ to be $false$, which violates the corresponding intent-compliant contract.
Therefore, we force the simulation to obey the contract and associate the corresponding route with the Boolean condition of this behavior (c1). As a result, $B$
receives and selects $[B, C, D]$ as its path, then sends it to its neighbors.
In the next step, when $F$ receives route
$[F, A, B, C, D]$ and compares it with $[F, E, D]$, $F$'s configuration would set $isPreferred$ for this received route to be $true$. However, the intent-compliant contracts require $F$ sets $isPreferred$ for the two paths comparison to be $true$, contradicting the original configuration. We again
force the simulation to follow the intent-compliant contracts. As a result,
$F$ keeps $[F, E, D]$ as its path, instead of $[F, A, B, C, D]$, and also associate $[F, E, D]$ with the corresponding condition (c2). 
In the end, each
router selects the path underlined in the Figure.

\para{Step 3: Map violated contracts to configuration errors.}
For each violated contract, \system{} systematically maps it to its corresponding configuration snippet based on the type of violation and details such as routes and devices involved, as shown in Table~\ref{tab: contracts}.
In our example, two contracts are violated. 
In the first case, $C$ sets $isExported$ of $[C, D]$ to $false$. We map this violated contract to the export policy snippets on $C$ for neighbor $B$, specifically the $filter$ policy (lines 3-5), which matches $[C, D]$ with the prefix filter and incorrectly discards it.
In the second case, $F$ sets $isPreferred$ of $[F, A, B, C, D]$ over $[F, E, D]$ to be $true$. We map this violated contract to the import policies on $F$ that match $[F, A, B, C, D]$ and $[F, E, D]$, the $setLP$ policy of $F$ (lines 3-9), where the first rule matches the AS-path of $[F, A, B, C, D]$ and sets the local preference to 200, while the second rule matches $[F, E, D]$ and sets its local preference to 80.
Both snippets are consistent with the ground truth presented at the beginning of our example (\S\ref{sec:limitations}).

\para{Step 4: Generate conflict-free repair patches with contract-specific template-based constraint programming.}
Given a set of localized configuration errors, a naive repair approach is to recompute an intent-compliant configuration from the erroneous one using constraint programming (\eg, network synthesis~\cite{el2018netcomplete}). Specifically, we model the configuration as constraint $C$, where the parameters of erroneous snippets are symbolic, the violated contracts to fix as constraint $V$, and the non-violated contracts to preserve as constraint $P$. The repair patch computation is then expressed as finding a satisfiable assignment for $C \land V \land P$.

However, this approach has two main limitations: (1) encoding all configurations and contracts together leads to high computational cost, and, (2) contract conflicts on the same configuration snippet may cause unsatisfiable solutions. For example, if two contracts for different prefixes map to the same policy rule, with one requiring local preference >100 and the other <100, no single assignment can satisfy both, making the repair infeasible without changing the rule structure. To address these limitations, we introduce the \textit{contract-specific template}, which uses fine-grained policy matching by adding a new rule that matches specific attributes of the route in the contract, such as prefix, community, and AS-path. This ensures that the rule only matches the intended route and does not affect others. The corresponding action is modeled as a symbolic variable, which is solved through constraint programming. This approach allows the repair problem to be decomposed into independent subproblems, and conflicts between contracts on the same configuration snippet are resolved, ensuring a satisfiable solution.

In this example, the contract-specific template for $isExported$ of route $[C,D]$ inserts a new match rule before the currently matching one that uniquely matches this route. The action is modeled as a symbolic variable, which is solved through constraint programming to set the permit action as $True$. Likewise, for $isPreferred$ of routes $[F,A,B,C,D]$ and $[F,E,D]$, the template inserts a rule that matches $[F,A,B,C,D]$ and sets the permit action as $True$ and the local-preference to a value less than that of $[F,E,D]$ ($<80$).

\section{Basic Design}\label{sec:basic}
We now elaborate the basic design of \system{} with more details. For ease of
presentation, we first focus on single destination prefix and   
describe how \system{} computes an intent-compliant data plane and derive the
corresponding contracts (\S\ref{sec:derive-contracts}), and how it identifies
all contract violations in the configurations using symbolic simulation and
repairs them (\S\ref{sec:locate-and-repair}). We then introduce how \system{}
supports complex configurations such as route aggregation, ACL, multi-path
routing (\S\ref{sec:complex}).

\para{Intents.} Figure~\ref{figure: intent syntax} illustrates the syntax of the language \system{} uses to specify intents. Specifically, each intent is represented as a $(identifier,\ path\_req)$ tuple, where $identifier$ specifies the source and destination routers, and the $path\_req$ encodes the path requirement. A path requirement consists of a regular expression over devices ($path\_regex$), a type specifier (\textbf{any} or \textbf{equal}), and a failure scenario specifier that requires the intent to hold under up to arbitrary $K$ link failures.

This syntax is expressive enough to capture common intents such as reachability, waypoint, and multi-path reachability. For example, in Figure~\ref{fig:example}, the reachability intent from router $B$ to router $D$ is specified as $(B.^*D, \mathbf{any}, failures=0)$, and the waypoint intent through router $C$ is specified as $(A.^*C.^*D,\mathbf{any}, failures=0)$.
A multi-path reachability intent can be specified as $(S.^*D, \mathbf{equal})$, indicating that all available paths from $S$ to $D$ should be used equally.




\begin{figure}[t]
\setlength{\abovecaptionskip}{0em}
\setlength{\belowcaptionskip}{0em}
\begin{tabular}{rcl}

  $ints$& $::=$ & $int^*$ \\
  $int$ & $::=$ & $(identifier,\hspace{0.5em} path\_req)$\\
  $identifier$  & $::=$ & $(srcDev,\hspace{0.5em} srcIp,\hspace{0.5em} dstDev,\hspace{0.5em} dstIp)$\\
  $path\_req$ & $::=$ & $(path\_regex,\hspace{0.5em} type,\hspace{0.5em} failures = K$) \\
  $type$ & $::=$ & $\mathbf{any} \hspace{0.5em} | \hspace{0.5em} \mathbf{equal}$ \\    

\end{tabular}
\caption{The syntax of \system{}’s intent language.}
\label{figure: intent syntax}
\vspace{-10pt}
\end{figure}

\vspace{1em}
\subsection{Intent-Compliant Contracts}\label{sec:derive-contracts}

Intent-compliant contracts are predicates on router behaviors such that, if satisfied, the network will yield an
intent-compliant data plane. Given a set of intents and an erroneous
configuration, we find such contracts in two steps: (1) compute
an intent-compliant data plane with small differences from the data plane generated by the erroneous configuration; and (2) derive intent-compliant
contracts accordingly.

\para{Compute an intent-compliant data plane by finding paths for unsatisfied intents.}
Suppose the data plane generated by the erroneous configuration satisfies $m$
intents and violates $n$ intents.  We use the valid paths in the erroneous data
plane for those $m$ intents as an initial set of path constraints. For each of
the $n$ unsatisfied intents, we find it a shortest valid path overlapping with
existing constraints as mush as possible (\ie, preferably having a
superpath/subpath relation) and without breaking the existing constraints (\ie,
the founded path must not form any loop with paths in the path constraints), and
add this path into the set of path constraints.  We find such a valid path using deterministic finite automaton (DFA)
multiplication. During the iteration, given an intent $x$, if no path satisfying
$x$ and the path constraints can be found, we gradually remove paths from the
path constraints one at a time, mark the corresponding intents as unsatisfied,
and try to find a valid path for $x$ satisfying the new set of path constraints.
This backtracking continues until such a path is found and then we move on to
the next unsatisfied intent. 

To make sure we can quickly find an intent-compliant data plane, we cope with
the combinatorial complexity of the order of path finding for unsatisfied intents and
the order of path constraints backtracking with the following two principles:

\para{Path finding for unsatisfied intents: more constrained intents first and
recently backtracked intents first.} 
To be concrete, more constrained intents refer to ones that require not only
reachability, but sequence of routers in paths (\eg, waypointing). We try to
find shortest valid paths for them with priority because their sets of
intent-compliant paths are typically smaller than those of intents on only
reachability, and more likely become empty when there are too many path
constraints to satisfy simultaneously. For example, suppose none of the intents in the example network (Figure~\ref{fig:example}) are satisfied, we prioritize computing paths for $A$ and $F$'s intents over all other routers' intents.
Next,
recently backtracked intents refer to ones whose paths are recently removed from
the path constraints to increase the chance for finding a valid path for an
unsatisfied intent. We find paths for them with priority because their sets of
intent-compliant paths just shrank due to the recent removal of their valid
paths. 

\para{Path constraints backtracking: closest path first and newest added
path first.} 
Supposed for intent $x$, we could not find an intent-compliant path without
breaking the existing path constraints. We examine the paths in the path
constraints and remove the one whose source router is closest, in terms of hop
count, to the source router of $x$. 
For example, when backtracking to compute $A$'s intent, we first remove the closer paths of $B$ and $F$ instead of the farther ones like $C$ or $E$. 
Our rationale is that by removing such a closest path, we have a better chance for
finding $x$ a valid path. Next, the concept of newest added path is self
explanatory. 
By removing the newest added path, we seek a valid path for $x$ without causing too many changes to the set of existing valid paths.

Although these principles are not perfect in theory and may require computing intent-compliant paths in factorial orders of intents in the worst case. In practice, the process efficiently computes an intent-compliant data plane for real and synthetic networks of various scales (\S\ref{sec:eval}).

\para{Derive intent-compliant contracts via path existence conditions.}
An intent-compliant data plane can be decomposed into a set of
intent-compliant contracts of routers via the sufficient and necessary
conditions of path existence. Specifically, given a path $R_1 \rightarrow
\ldots, \rightarrow R_{i-1} \rightarrow R_i \rightarrow R_{i+1}, \ldots,
\rightarrow R_n]$, it exists in the data plane if and only if for each router
$R_{i}$, it peers with $R_{i+1}$, accepts and prefers $R_{i+1}, \rightarrow,
\ldots, \rightarrow, R_{n}$ over all other routes, and announces $R_{i},
\rightarrow, R_{i+1}, \ldots, \rightarrow, R_{n}$ to $R_{i-1}$. Thus, to ensure a path in the intent-compliant data plane exists, each router on the path must satisfy contracts related to peering, route selection, and announcement behaviors.

\vspace{1em}
\subsection{Diagnose and Repair Errors} \label{sec:locate-and-repair}


With the intent-compliant contracts derived from the data plane, \system{} repairs configuration errors in three steps: (1) identifying all violated contracts using selective symbolic simulation; (2)
localizing errors by mapping violated contracts to configuration snippets; (3) computing conflict-free repair patches using template-based constraint programming.

\para{Identify violated contracts using selective symbolic simulation.} 
Given the set of intent-compliant contracts and the original configuration, \system{} simulates the protocol behavior. At each step, it checks whether the router’s behavior—such as importing, exporting, preferring, or peering—matches the corresponding contract. If a behavior violates the expected contract, \system{} forces the behavior to conform to the contract and selectively switches to the symbolic variant of the configuration where the violation has been corrected. The violated contract is then attached to the corresponding route as an annotation. This selective symbolic simulation proceeds iteratively until the resulting data plane converges to the intent-compliant one. In the end, \system{} collects all encountered violated contracts for error localization and repair.

Although most configurations can be handled in a prefix-independent manner for path vector protocols, which enables independent symbolic simulation for each prefix, the establishment of a peering session for one prefix can affect the propagation of all prefixes, as sessions are established at the router level. Therefore, \system{} treats the $\mathit{isPeered}$ contract as shared and enforces it across all prefixes during symbolic simulation: if any prefix requires a peering session on a given link, \system{} forces all prefixes to establish that peer relationship.


\para{Localize configuration errors by mapping violated contracts to configuration snippets.}
Given all violated contracts identified through selective symbolic simulation, along with details such as routes and devices, \system{} maps each violation to its corresponding erroneous configuration snippet (as shown in Table \ref{tab: contracts}).
For example, a violated $\mathit{isImported}$ contract on device $u$, involving neighbor $v$ and route $r$ with attributes such as AS-path, community, and next-hop, is mapped to the import policy rule on $u$ that matches $r$ from $v$.
Similarly, a violated $\mathit{isPeered}$ contract is mapped to the relevant neighbor statements and associated interface configurations on the involved devices. 
This fine-grained mapping pinpoints the exact configuration locations needing repair.


\para{Repair configuration errors using contract-specific template-based constraint programming.}
Given all localized configuration errors, the direct approach to repair them is modeling the repair process as a constraint-solving problem. Specifically, the parameters in the erroneous configuration snippets are symbolized, forming a partially specified configuration, denoted as $C$. Let $V$ represent the violated contracts to be fixed and $P$ represent the non-violated contracts to be preserved. The repair problem is encoded as a constraint problem of finding a satisfiable assignment for $C \land V \land P$, aiming to find parameter assignments that satisfy all contracts.

However, naively encoding all contracts and configurations together is computationally expensive, and may lead to unsatisfiable assignments due to conflicts between contracts for different prefixes mapped to the same configuration snippet.
For policy-related contracts (e.g., $\mathit{isImported}$, $\mathit{isExported}$, $\mathit{isPreferred}$), \system{} utilizes contract-specific templates which leverage BGP’s fine-grained policy control to exactly match the route. 
We summarize the templates used for each contract in Appendix~\ref{appendix: templates}.
This involves adding a rule that matches the specific route before the current matched one, ensuring the patch only affects the desired route. Such a design allows independent repair of each contract, resolving conflicts between contracts for different prefixes on the same configuration snippet and ensuring a solution. Constraint programming is then used to find a satisfiable value for the action parameter to correct the contract.
For the $\mathit{isPeered}$ contract, \system{} resolves dependencies during symbolic simulation. Patches can be computed independently using a template that models the peer configuration, with constraint programming finding the necessary assignments to establish the peer relationship.

\vspace{1em}
\subsection{Support Complex Configuration}\label{sec:complex}
To support complex configurations, \system{} extends mechanisms for route aggregation, ACLs, and ECMP.

\para{Route aggregation.}
Route aggregation combines multiple IP prefixes into a single, larger prefix to reduce routing table size and improve efficiency. The main challenge is ensuring the aggregated prefix's propagation path satisfies all contracts from its sub-prefixes. To address this, \system{} solves the contracts collectively, rather than independently for each sub-prefix. By considering all contracts together and using template-based repair for the involved prefixes, \system{} creates a repair patch that consistently satisfies all contracts, avoiding discrepancies in the aggregated route’s propagation path. If no solution is found, \system{} employs a disaggregation strategy, splitting the aggregated prefix into its original components so that each prefix’s contract can be satisfied individually.

\para{Access control list.} 
The ACL allows a router to block traffic in the data plane. To support this, \system{} introduces two additional contracts, $\mathit{isForwardedIn}$ and $\mathit{isForwardedOut}$ (as shown in Table~\ref{tab: contracts}), which specify whether packets destined for a given prefix are allowed to enter or leave the router along the intended forwarding paths. During symbolic simulation, \system{} checks these contracts by comparing ACL behavior with the intent-compliant data plane. If a packet that should be forwarded is blocked by an ACL, the corresponding contract is marked as violated.
Due to the fine-grained nature of ACL rules, such violations can be diagnosed and independently repaired using contract-specific templates within the constraint-solving framework.


\para{Multi-path routing.} 
Multi-path routing is a strategy where packet forwarding to a single destination occurs over multiple equal-priority paths (ECMP).
To support this, \system{} introduces two extensions. First, when computing the intent-compliant data plane, it records all equally preferred paths, allowing a node to have multiple next hops and ensuring ECMP semantics are preserved. Second, \system{} introduces a new contract, $\mathit{isEqPreferred}$, specifying two routes be equally preferred at a given node.
When this contract is violated, \system{} identifies the configuration error and repairs it independently, similar to how $\mathit{isPreferred}$ contracts are handled.

\begin{figure}[tbp]
  \centering
  \begin{subfigure}[b]{0.45\textwidth}
    \includegraphics[width=\linewidth]{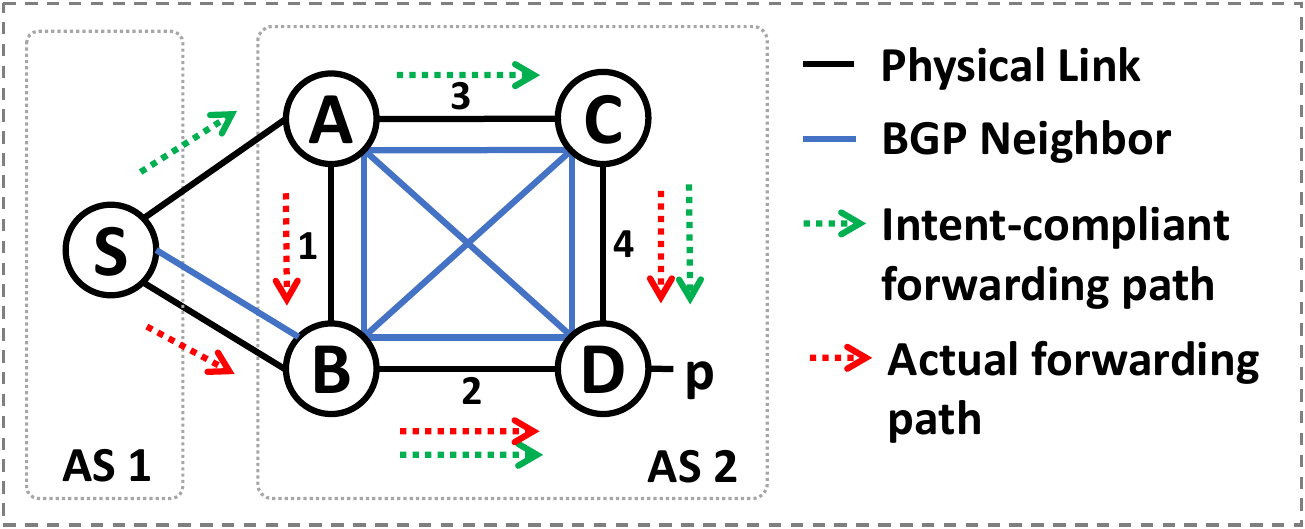}
    \caption{Intent-compliant data plane.}
    \label{fig:multiproc-DP}
  \end{subfigure}
   \hspace{2em}
  \begin{subfigure}[b]{0.45\textwidth}
    \includegraphics[width=\linewidth]{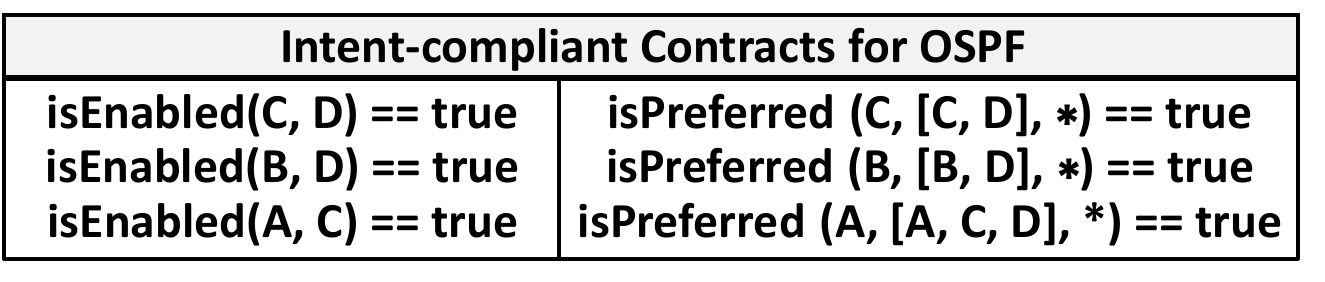}
    \vspace{-14pt}
    \caption{Intent-compliant contracts for the OSPF network.}
    \label{fig:multiproc-OSPF}
  \end{subfigure}
  \caption{The multi-protocol example network.}
  \vspace{-20pt}
  \label{fig:multiproc}
\end{figure}

\vspace{-10pt}
\section{Multi-Protocol Networks}\label{sec:multi-protocol}
\vspace{-10pt}

After describing \system{}'s basic design using BGP as an example, we introduce
how it diagnoses and repairs networks where multiple protocols (\eg, OSPF and
BGP) co-exist. Because \system{} can handle a network running different routing
protocols in different regions by examining these regions separately, we focus
on networks running different routing protocols in a layered setting (\eg, OSPF
as underlay and BGP as overlay). Error diagnosis and repair for such networks is more challenging due to the dependency between these protocols.

To this end, we present our solution for multi-protocol networks in \S\ref{subsec:hybrid}, and describe how to extend the selective symbolic simulation approach to diagnose and repair link-state routing protocols in \S\ref{subsec:ospf}. 


\vspace{1em}
\subsection{Our Solution}
\label{subsec:hybrid}


Inspired by modular verification approaches such as Kirigami~\cite{thijm2024kirigami} and Lightyear~\cite{tang2023lightyear}, we use an assume-guarantee approach, and decompose the intents of the multi-protocol network into distinct layers. 
The approach first assumes the proper functioning of the underlay network
(e.g., reachability between overlay-neighboring routers),
and diagnosis and repairs the overlay network to be intent-compliant.
It then uses the assumption as the intents for the
underlay network and repeat the diagnosis and repair process.
We use the following example to illustrate our key ideas.

\para{Example.}
Figure~\ref{fig:multiproc-DP} shows an example of two ASes peered with eBGP and routers $A$, $B$, $C$, and $D$ are connected using OSPF in underlay and peered with iBGP in a full mesh in overlay. In the configuration, $S$ is only peered with $B$ via eBGP and the OSPF cost of each link in $AS$ 2 is shown in the edges. The destination IP prefix $p$ is located at node $D$ and is advertised via BGP.
The network intents are: (1) all routers can reach prefix $p$, and (2) the path of $S$ to reach $p$ should not pass $B$. The data plane of the original configuration is marked in red, which is erroneous. The actual forwarding path [S, B, D] violates Intent 2. This is due to two configuration errors: first, $S$ lacks a BGP peer with $A$; second, misconfigured OSPF costs in AS 2 cause $A$ to prefer reaching $D$ via $B$ instead of $C$.

\para{Derive intents for overlay and underlay.} We begin by verifying the specified intents against the generated data plane, classifying them as either satisfied or violated. These intents are then decomposed into overlay and underlay intents.

In this example, Intent (1) is satisfied and corresponds to the overlay. Intent (2) is violated and must be split into overlay and underlay components. To satisfy it, we compute the shortest valid path in the physical topology and determine that $S$ must reach $D$ via $[S, A, C, D]$. Based on the domain knowledge that iBGP nodes do not re-advertise routes learned from other iBGP peers (i.e., path $[A, C, D]$ cannot be learned via iBGP), and noting that $A$, $C$, and $D$ belong to the same IGP domain, we derive the following sub-intents: 1) BGP Intent 1: $S$ reaches $D$ via $[S,A,\cdots,D]$, and 2) OSPF Intent 1: $A$ reaches $D$ via the path $[A,C,D]$. 
Additionally, since BGP relies on OSPF to establish neighbor sessions using OSPF-learned IPs, even between non-directly connected nodes (e.g., $A$ and $D$), we derive OSPF Intent 2: $D$ and its neighboring routers $A$, $B$, and $C$ must be mutually reachable.



\para{Process each layer independently.} 
Given the derived BGP intents, we generate the BGP intent-compliant data plane, and selectively simulate the BGP network to find violated contracts. In this example, we identify a violated $isPeered$ contract between $S$ and $A$, and generate a repair that establishes the missing peer.
Similarly, using the derived OSPF intents, we build the OSPF intent-compliant data plane, and then apply selective symbolic simulation to diagnose and repair the underlay network, as provided in \S\ref{subsec:ospf}. 

\vspace{1em}
\subsection{Link-State Routing Protocols}\label{subsec:ospf}
\vspace{0.5em}

The fundamental difference between path-vector and link-state protocols is that the former selects the best paths for different prefixes independently (\ie,
per-prefix policy routing) while the latter does so with a unified metric (\ie,
prefix-oblivious link costs). Thus, we must first collect all OSPF intents across different prefixes before deriving the intent-compliant contracts. Then, we follow the design of path-vector protocols in \S\ref{sec:basic} to handle link-state routing protocols.





\para{Define contracts.} Unlike BGP, OSPF does not establish peers but instead relies on OSPF-enabled interfaces. Additionally, OSPF does not support route policies and determines routing paths solely based on link costs. Accordingly, we define two types of contracts for OSPF: 1) $isEnabled(a,b)$: the interfaces between nodes $a$ and $b$ are OSPF-enabled, and 2) $isPreferred(u,r,r')$:  identical to the BGP definition, it specifies that router $u$ should prefer route $r$ over $r'$.

\para{Derive intent-compliant data plane, and perform selective symbolic simulation to find violated contracts.} Given the set of underlay intents, we compute the corresponding intent-compliant data plane and derive the associated contracts.
Figure~\ref{fig:multiproc-OSPF} illustrates the contracts for OSPF in Figure~\ref{fig:multiproc-DP}, where $isEnabled$ contracts are derived from OSPF Intent 2, and $isPreferred$ contracts are inferred from the required forwarding behavior at nodes $A$, $B$ and $C$.

To reuse the simulation method to diagnose configuration errors, we simulate OSPF using a path-vector abstraction, where path selection is based on the cumulative cost.
In this example, the diagnosis reveals a contract violation at node $A$, which incorrectly prefers the path $[A,B,D]$ over $[A,C,D]$, indicating that the OSPF cost configuration is incorrect.



\para{Model the repair process as a MaxSMT problem.}
In OSPF networks, contract violations include enabled and preference errors.
OSPF-enabled error (\ie, violations of $isEnabled$ contracts) can be resolved by enabling OSPF on the relevant interfaces.
Preference error (\ie, violations of  $isPreferred$ contracts), however, requires adjusting link costs. Since OSPF computes a single forwarding tree per destination, modifying link costs for one path may inadvertently affect the forwarding behavior of other nodes.

To address the issue, we adopt the constraint programming approach to carefully adjust link costs while preserving as much of the original configuration (costs) as possible.
Specifically, we encode the OSPF network and its associated contracts into a MaxSMT formulation.
Hard constraints ensure that violated contracts are repaired and that non-violated contracts remain satisfied. 
Soft constraints aim to preserve the original link costs to minimize configuration changes.
In this example, we encode the repair problem as: 
\vspace{-0.5em}
\begin{flalign*}
  &\ (hard) \quad \{l_{CA} + l_{AB}+ l_{BD}  > l_{CD}\} \land  \{l_{BA} + 
  l_{AC}+ l_{CD}  > l_{BD}\} \\
  &\ \hphantom{(hard) \quad}\land \{l_{AB} + l_{BD} > l_{AC}+ l_{CD}\}  \\ 
  &\ (soft) \quad l_{AB}==1\land l_{BD}==2 \land l_{AC}==3 \land l_{CD}==4
\end{flalign*}
where $l_{a,b}$ denotes the link cost between nodes $a$ and $b$. 
In the proposed example, a valid repair solution is to update $l_{AB}$ to $7$, while keeping all other link costs unchanged. 

\section{Tolerating K-link Failure}
\label{sec:k-failure}

This section presents our designs to provide fault-tolerant capability. We focus on the $k$-link failure reachability, which refers to the network's ability to preserve connectivity between a given source and destination node pair in the presence of up to \(k\) arbitrary link failures.


\vspace{1em}
\subsection{Key Idea}
\label{subsec: 7.1}


Our key idea is to construct a fault-tolerant data plane where at least one path exists for the forwarding traffic under arbitrary \(k\) link failure(s). 
Specifically, we compute $(k+1)$ edge-disjoint paths for each \(k\)-link failure tolerance intent. Two edge-disjoint paths mean that the edges of them do not intersect. Since these paths are edge-disjoint, each link failure results in, at most, one path failing to forward the packet. According to the Pigeonhole Principle, at least one path must exist to forward the packet when $k$ links fail.

For multi-protocol networks, we use a similar approach in \S\ref{sec:multi-protocol} to decouple the derived fault-tolerant data plane into overlay intent-compliant data plane and underlay intent-compliant data plane. Then we derive intent-compliant contracts and perform selective symbolic simulation to diagnose and repair errors for each layer separately.




\begin{figure}[htbp]
  \centering
  \begin{subfigure}[t]{0.44\linewidth}
    \includegraphics[width=\linewidth]{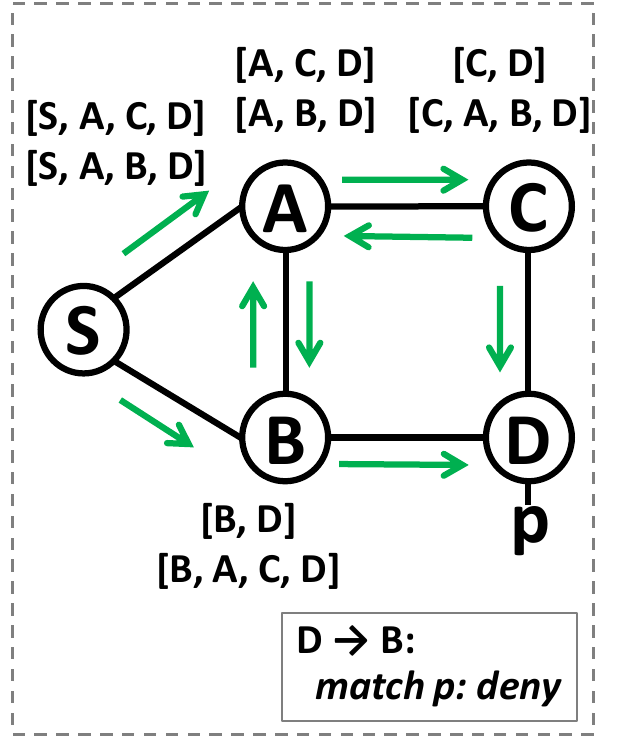}
    \caption{The intent-compliant fault-tolerant data plane.}
    \label{fig:failure-dp}
  \end{subfigure}%
  \begin{subfigure}[t]{0.5\linewidth}
    \includegraphics[width=\linewidth]{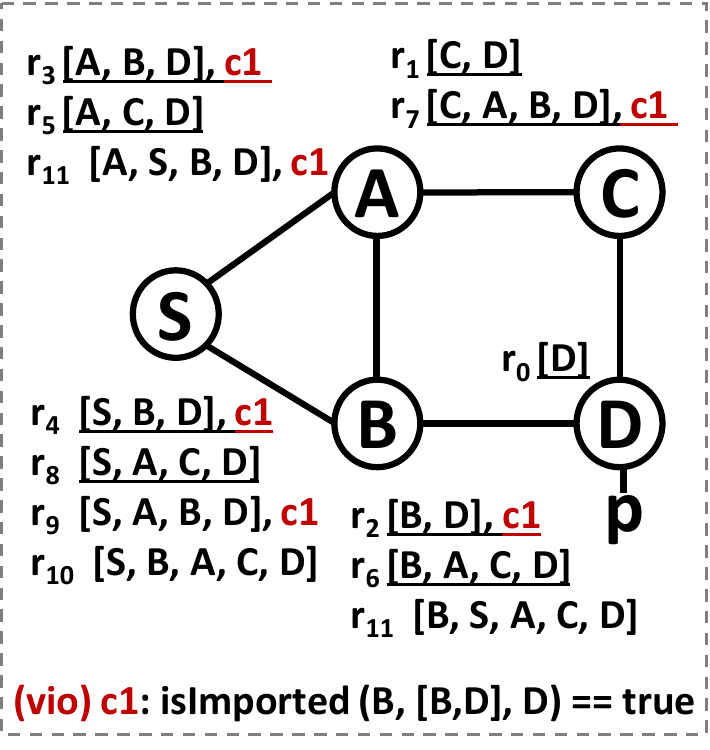}
    \caption{The symbolic simulation with single-link failures.}
    \label{fig:failure-sim}
  \end{subfigure}
  \caption{The single-link failure tolerance example network.}
  \vspace{-20pt}
  \label{fig:failure}
\end{figure}

\vspace{1em}
\subsection{Design Details}

As discussed in Section~\ref{subsec:ospf}, a link-state routing protocol can be simulated as a path-vector protocol, with the key difference being that link-state protocols compare only path costs and do not support routing policies. Therefore, we focus on BGP to illustrate the design details. 

\para{Example.}
Figure~\ref{fig:failure} illustrates an example network to ensure single-link failure tolerance, where the intent is that all routers must be able to reach prefix $p$ under any single-link failure.
In this network, five routers are connected via eBGP, all using the default BGP configuration except for router $B$, which is configured to drop routes from neighbor $D$ that match prefix $p$.
The intent is satisfied under normal conditions and remains valid under certain single-link failures.
However, for failures such as $(C,D)$ or $(A,C)$, since $B$ drops $D$'s route, this breaks the reachability intent under these failure scenarios. 

\para{Derive fault-tolerant data plane and contracts.}
For each $k$‑link failure tolerance intent, we compute $k+1$ edge‑disjoint paths and merge them into a fault‑tolerant data plane. These $k+1$ paths are obtained iteratively by invoking the shortest path algorithm $k+1$ times, each time removing the edges computed in the previous iteration from the graph.
Figure~\ref{fig:failure-dp} shows the merged fault-tolerant data plane of the example network with single-link failure tolerant, where each node is associated with two edge-disjoint paths, indicating that the traffic must be forwarded in either of the paths. For example, $[B,A,C,D]$ and $[B,D]$ are the derived data plane for $B$ to ensure reachability to $p$ even under single-link failure.

In the intent-compliant data plane where each node may have multiple forwarding paths, we derive the \emph{isPeered}, \emph{isExported} and \emph{isImported} contracts to ensure the existence of these paths, along with \emph{isPreferred} contracts to guarantee their selection over non-forwarding alternatives. However, we do not derive \emph{isPreferred} contracts that impose an order among the forwarding paths themselves, as the specific choice among them is irrelevant for satisfying the reachability intent. In this example, the most important intent-compliant contracts for node $B$ are that both $[B,D]$ and $[B,A,C,D]$ are preferred over all other available routes.


\para{Selective symbolic execution to diagnose and repair errors.}
According to the fault-tolerant contracts, we simulate the router to select and propagate multiple routes to diagnose and repair errors as in previous designs.
In this example, as Figure~\ref{fig:failure-sim} shows, when $B$ imports the route $[B,D]$ from $D$, we find that the \emph{isImported} contract for this route is violated, so we record this violation as an error and continue the simulation. 
After that, when $B$ imports the route $[B,A,C,D]$, we find it is intent-compliant and select both $[B,D]$ and $[B,A,C,D]$ as $B$'s best routes and send them to neighbors. 
In the end, $C$, $A$, and $S$ also select and send multiple routes that comply with their intents. Thus, we can simulate the network and diagnose the error under different failure scenarios.


\vspace{1em}
\subsection{Discussion}

Beyond reachability intents, we also handle other non-trivial intents (\eg, waypoint and avoidance) under arbitrary $k$-link failures, following the same approach used for reachability. For each intent, we compute multiple compliant paths and ensure that at least one remains available across all failure scenarios.

However, constructing a fault-tolerant data plane for multiple non-trivial intents may lead to path explosion. Each such intent requires computing $k+1$ edge-disjoint paths while simultaneously preserving the path constraints imposed by other intents. In some cases, this necessitates backtracking to exhaustively enumerate valid path combinations across intents to find a satisfiable solution.
Developing more efficient algorithms to compute fault-tolerant data planes is our future work.
Different from waypoint or avoidance intents, supporting multiple $k$-link failure reachability intents does not cause path explosion. This is because such intents are always handled last (as the principle one in \S\ref{sec:derive-contracts}), and their compliant paths do not break existing path constraints, thereby avoiding the need for backtracking.

In practice, users may require path preference under failures, where some paths are designated as primary and others as backups. The current design can be extended to support such path preference intents, given an ordered set of paths (\eg, $P_1 \gg \dots \gg P_x$). \system{} can first ensure these paths are available and then enforce the preferences by adding \textit{isPreferred} contracts to the intersection nodes along these paths.

\begin{table}[t]
    \centering
    \includegraphics[width=\linewidth]{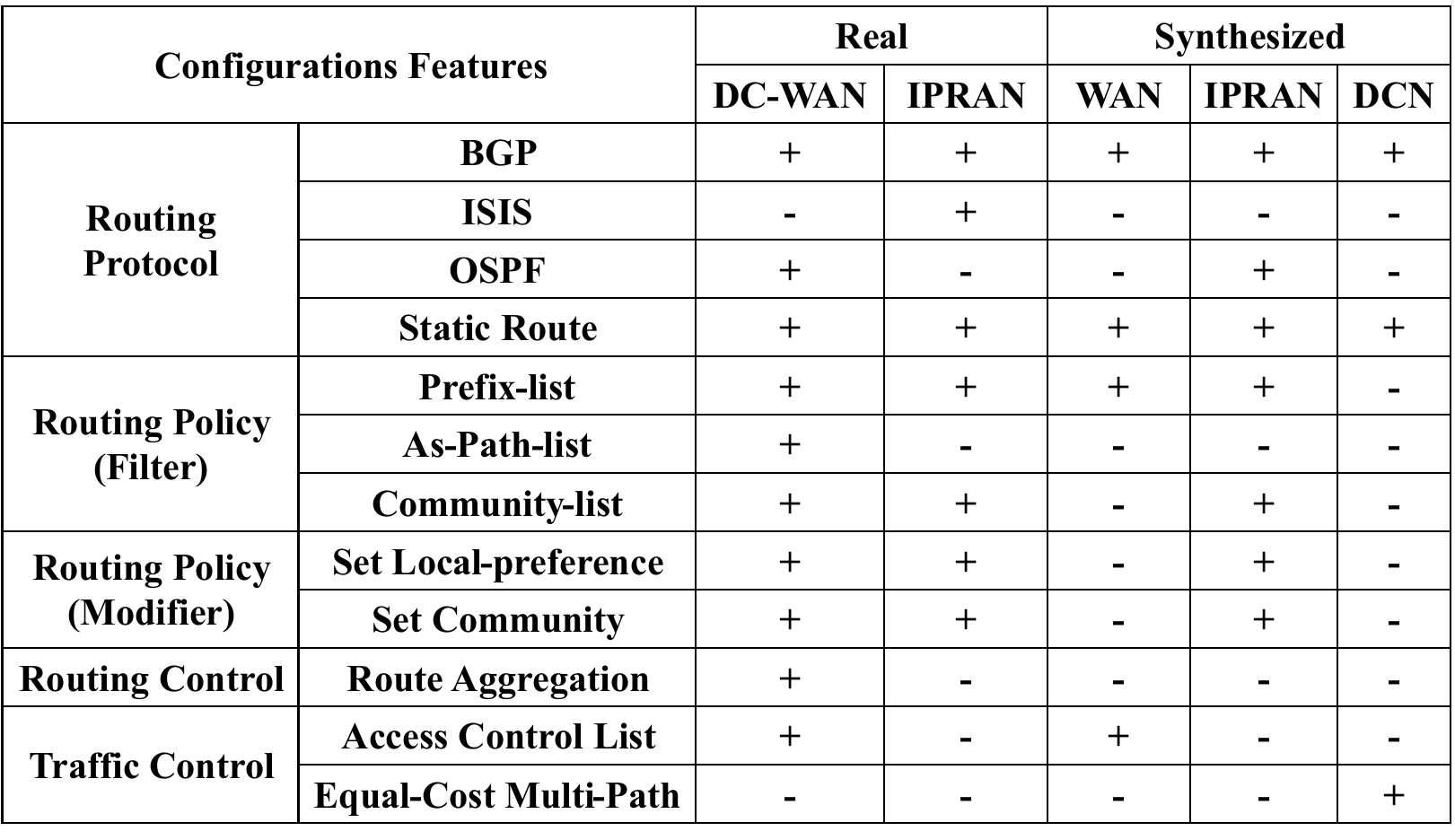}
    \caption{Configuration features of the evaluated networks. A + (–) indicates the feature is present (absent) in the network.} 
    \label{tab:real-config}
    \vspace{-10pt}
\end{table}

\begin{table*}[t]
    \centering
    \includegraphics[width=\linewidth]{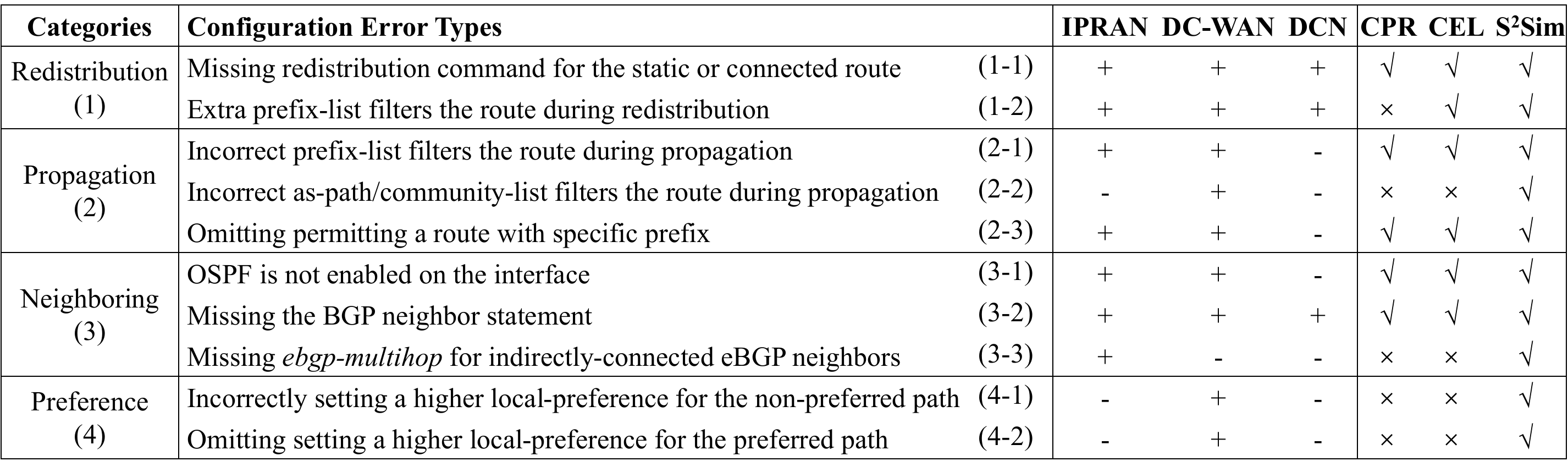}
    \caption{Errors introduced in synthesized configurations, all of which are observed in real networks. A + (–) indicates the presence (absence) of the error in the real network. A $\checkmark$ ($\times$) indicates whether the tool is able (unable) to handle the error.}
    \label{tab:real-error}
    \vspace{-10pt}
\end{table*}


\vspace{-10pt}
\section{Performance Evaluation}\label{sec:eval}
\vspace{-5pt}

We have implemented a prototype of \system{} in 8K LoC in Java as a plugin of Batfish\cite{fogel2015general}, and will open source it upon publication. 
All experiments were conducted on a Linux server equipped with two Intel Xeon Silver 4210R 2.40GHz CPUs and 128GB of DDR4 DRAM.

Our evaluation answers the following two questions: (1) Can \system{} and SOTA tools diagnose and repair real network configuration errors? (\S\ref{subsec:comparision}) (2) How efficiently \system{} diagnose and repair the errors for large-scale networks? (\S\ref{subsec: efficiency})



\para{Configurations.} We use both real network configurations and synthesized network configurations for evaluation. Table~\ref{tab:real-config} summarizes the configuration features of these networks.

\para{(1) Real network configurations.} 
One major public cloud provider offers an operating wide-area network (WAN) consisting of 88 nodes interconnecting data centers (i.e., \emph{DC-WAN}), with an average of 5K lines of configuration per node. Another major vendor provides four IP radio access network (i.e., \emph{IPRAN1}-\emph{IPRAN4}) with 36, 56, 76, 106 nodes, connecting base stations to base station controllers via routers, with an average of 300 lines of configuration per node. 
Both DC-WAN and IPRAN utilize the underlay-overlay architecture.
Additionally, the DC-WAN configuration provider also offers the configuration features of their operating DCNs. However, due to the large size of real DCN configurations, the corresponding configuration files are not available to us.

\para{(2) Synthesized network configurations.} We adopt the real WAN topologies from TopologyZoo (33 to 155 nodes), and use NetComplete \cite{el2018netcomplete} to generate WAN configurations with configuration lines 3K to 13K. We also consider two large-scale networks: IPRANs having nodes ranging from 1000 to 3000 and configuration lines from 176K to 561K, and data-center networks (DCN) using the fat-tree architecture, having nodes ranging from 80 to 1280 and configuration lines from 2K to 140K. We synthesize IPRAN/DCN configurations following the structure of real IPRAN/DCN networks.










\para{Intents.} 
We consider three types of intents for evaluation: (1) Reachability (RCH): a source node can reach a destination node without link failures, (2) Fault-tolerant reachability: a source node can reach a destination node with $k$-link failures, and (3) Waypoint reachability (WPT): a source node can reach a destination node via specific waypoint node(s).

\vspace{1em}
\subsection{Comparison with Existing Tools} \label{subsec:comparision}

We compare \system{} with CPR~\cite{gember2017automatically} and CEL~\cite{gember2022localizing} in both capability and running time. 
As discussed in Section~\ref{sec:limitations}, other tools such as Batfish, Minesweeper, ACR are not adequate for configuration diagnosis or repair. , which is why we select CPR (for repair) and CEL (for diagnosis) as our baselines.
However, since CPR and CEL lack support for certain features present in real configurations, we construct synthesized WAN configurations and inject errors supported by these tools but derived from real-world cases, to compare their running time.

\para{Errors in Real Configurations.}
Table~\ref{tab:real-error} summarizes common configuration errors observed in real DC-WAN and IPRAN networks. Additionally, the DC-WAN configuration provider has reported a list of typical errors found in their managed DCNs. 
We categorize these errors into four types. Redistribution-related and neighborhood-related errors often stem from missing configurations for route redistribution or the establishment of routing adjacency. These are the most frequently encountered issues across all real network configurations. Route propagation-related and preference-related errors typically result from incorrect or absent routing policies (e.g., filters), and are found only in DC-WAN and IPRAN networks. This is because these two network types rely heavily on policy-based routing for inter-domain traffic control, whereas DCNs generally do not require such mechanisms.

\begin{figure*}[htbp]
  \centering
  \begin{minipage}[t]{0.28\linewidth}
    \includegraphics[width=\linewidth, height=3cm]{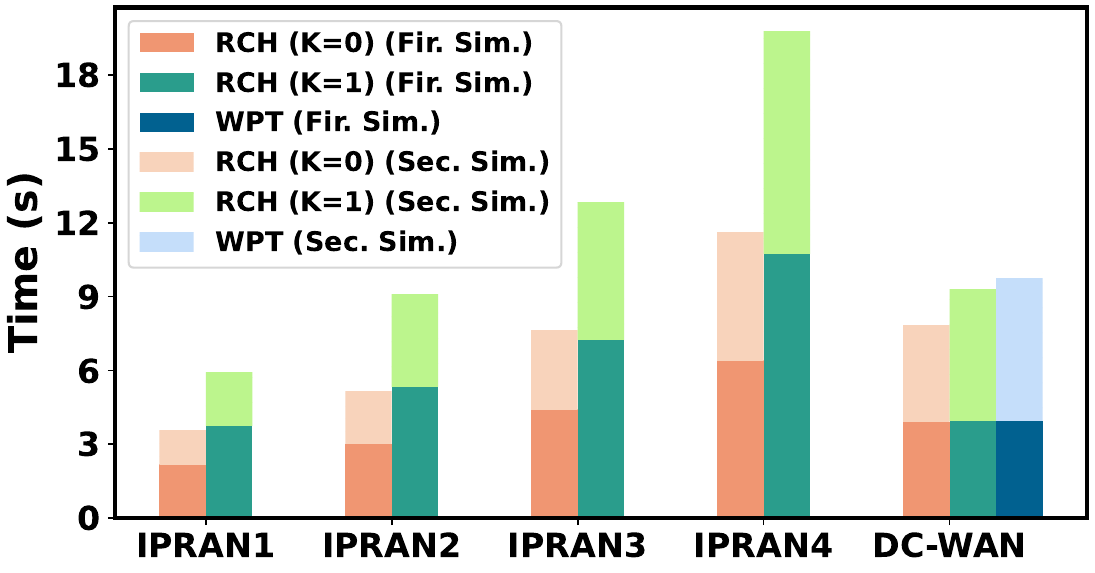}
    \caption{The runtime of \system{} on five real configurations, including two rounds of simulation.}
    \label{fig:real-config-exp}
  \end{minipage}
  \hspace{1mm}
  \begin{minipage}[t]{0.70\linewidth}
    \centering
    \begin{subfigure}[t]{0.49\linewidth}  
      \includegraphics[width=\linewidth, height=3cm]{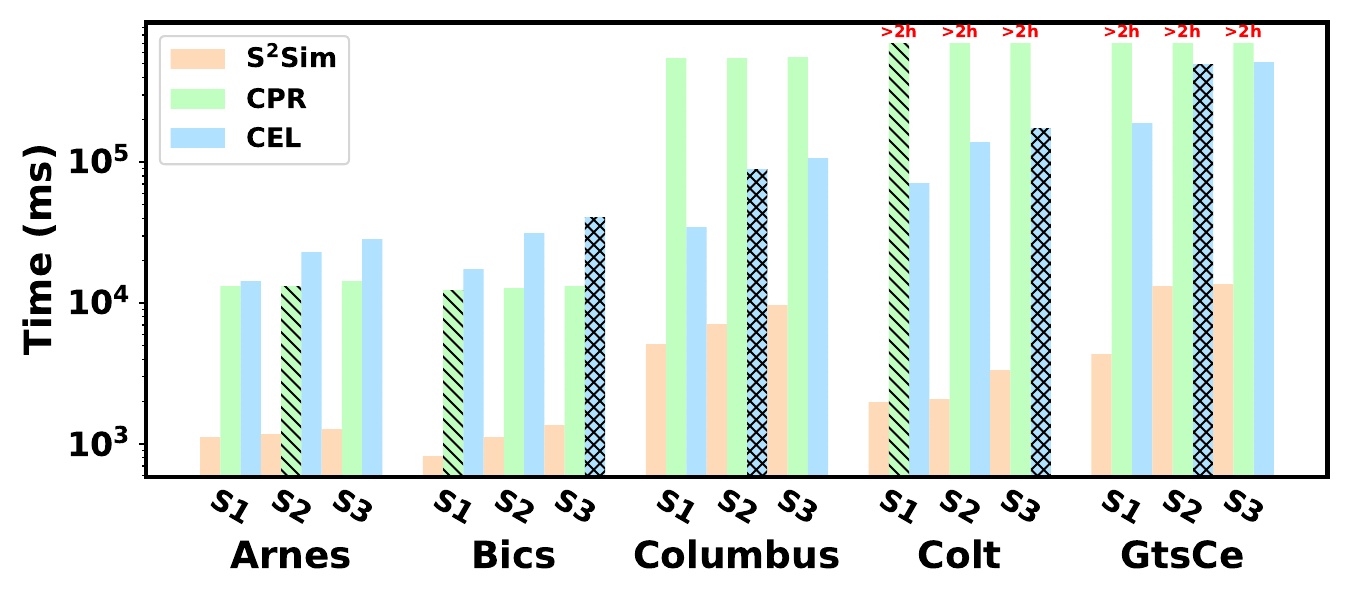}
      \caption{Reachability.}
      \label{fig:eval-init-a}
    \end{subfigure}
    \hfill
    \begin{subfigure}[t]{0.49\linewidth}  
      \includegraphics[width=\linewidth, height=3cm]{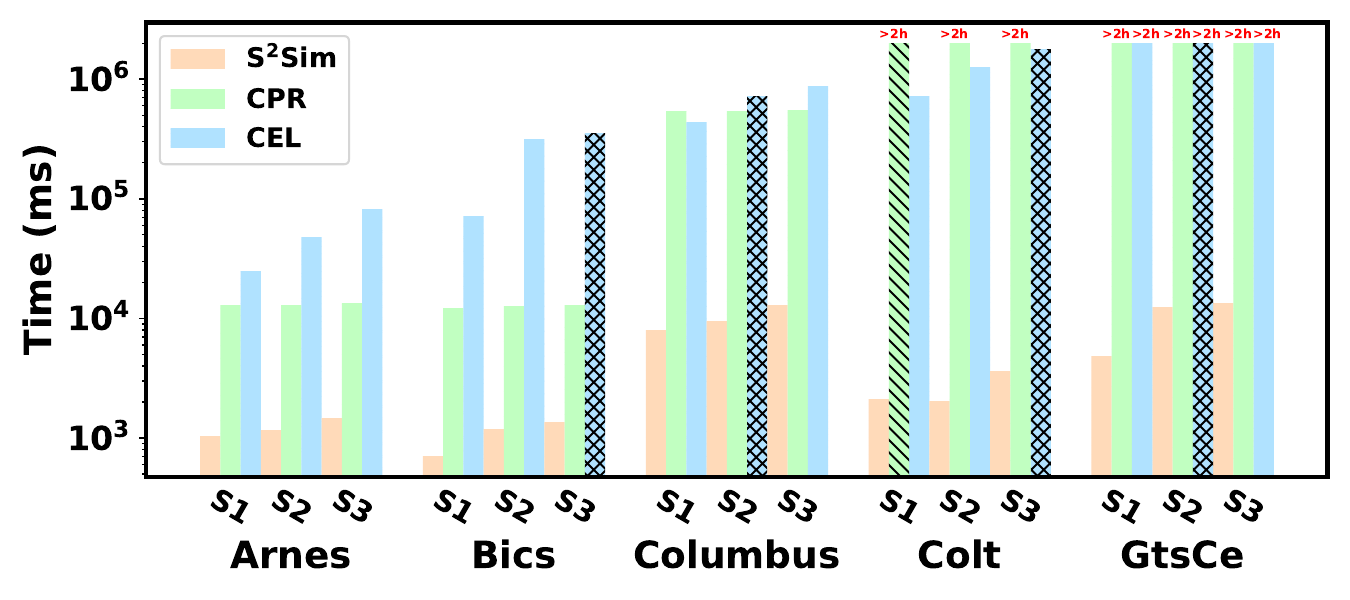}
      \caption{Fault-tolerant reachability (k=1).}
      \label{fig:eval-init-b}
    \end{subfigure}
    \caption{The runtime of \system{} and other tools on synthesized WAN configurations.}  
    \label{fig:syn-config-diff-tools}
  \end{minipage}
\end{figure*}

\begin{figure*}[htbp]
  \centering
  \begin{minipage}[t]{0.64\linewidth}
    \centering
    \begin{subfigure}[t]{0.49\linewidth}  
      \includegraphics[width=\linewidth, height=3cm]{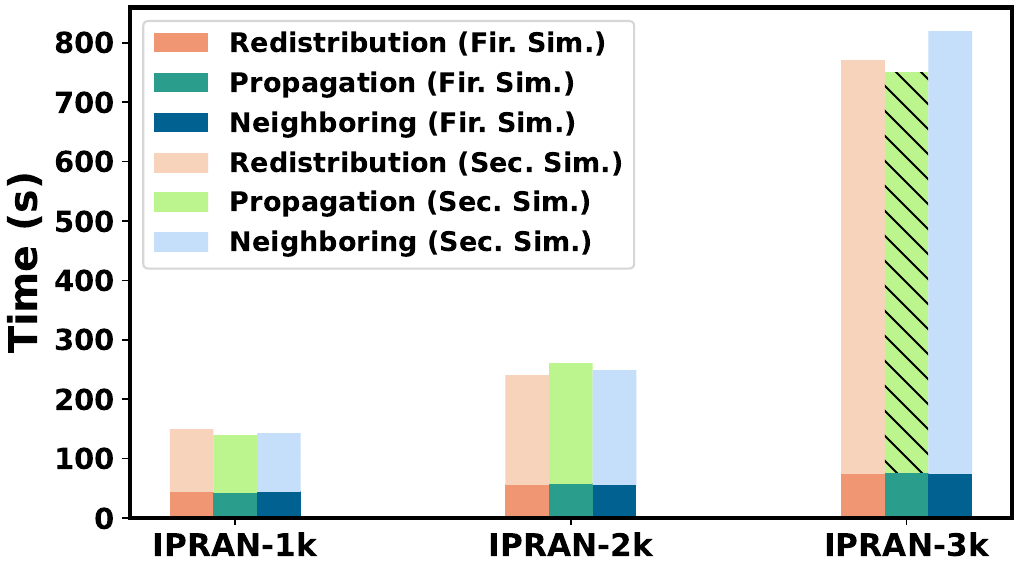}
      \caption{Error category vs. runtime.}
      \label{fig:impact-of-err-category}
    \end{subfigure}
    \hfill
    \begin{subfigure}[t]{0.49\linewidth}  
      \includegraphics[width=\linewidth, height=3cm]{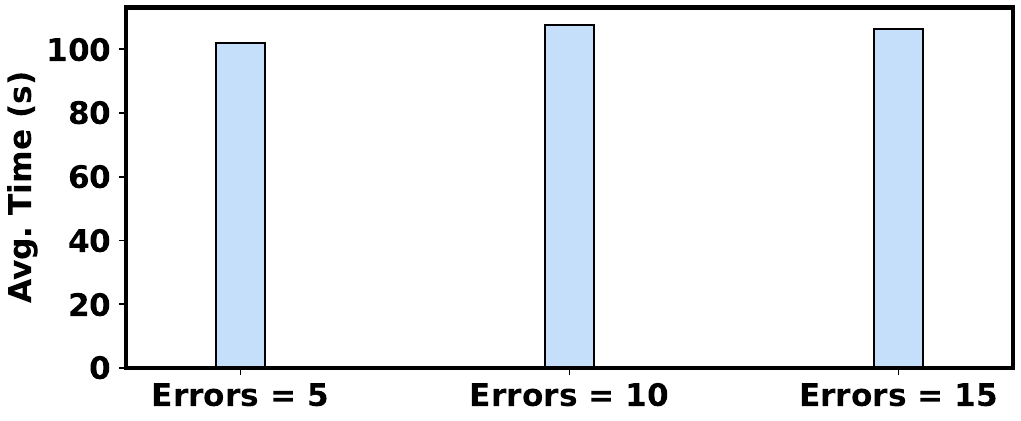}
      \caption{Error number vs. runtime.}
      \label{fig:impact-of-err-number}
    \end{subfigure}
    \vspace{-0.5em}
    \caption{Impact of error types and error number on \system{} runtime in IPRANs.}  
    \label{fig:impact-of-err}
  \end{minipage}
  \hspace{1mm}
  \begin{minipage}[t]{0.34\linewidth}
    \includegraphics[width=\linewidth, height=3.03cm]{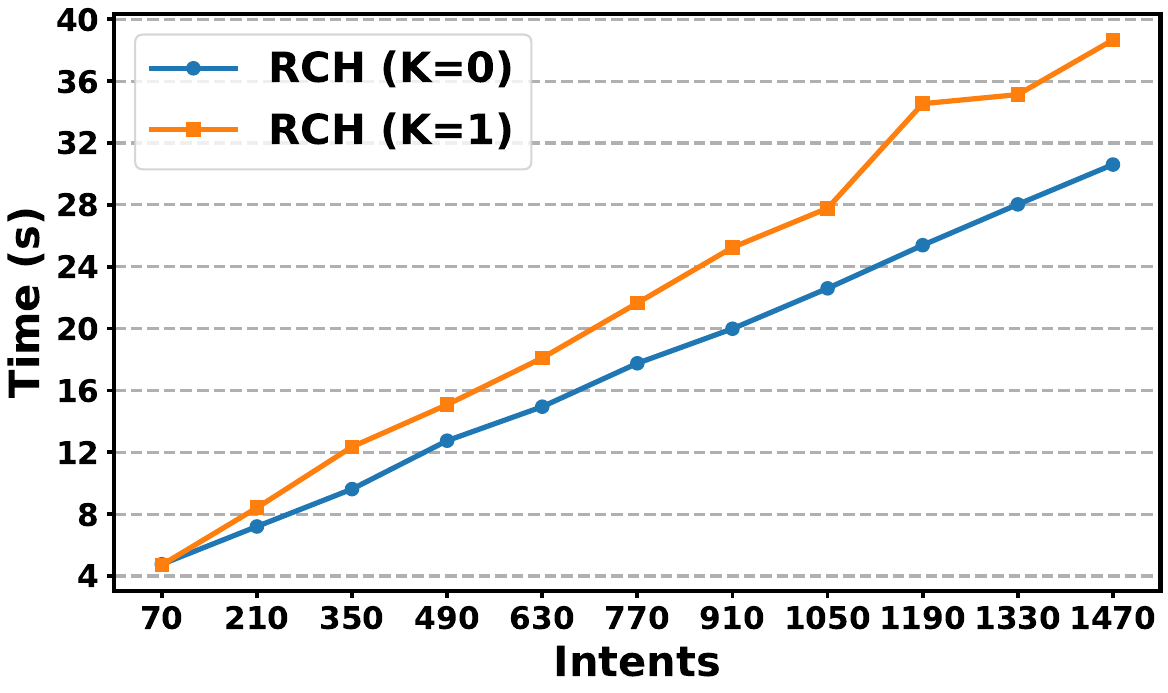}
    \caption{Impact of intent number on \system{} runtime in a DCN.}
    \label{fig:impact-of-intents}
  \end{minipage}
  \vspace{-10pt}
\end{figure*}

\begin{figure}
    \centering
    \includegraphics[width=0.85\linewidth]{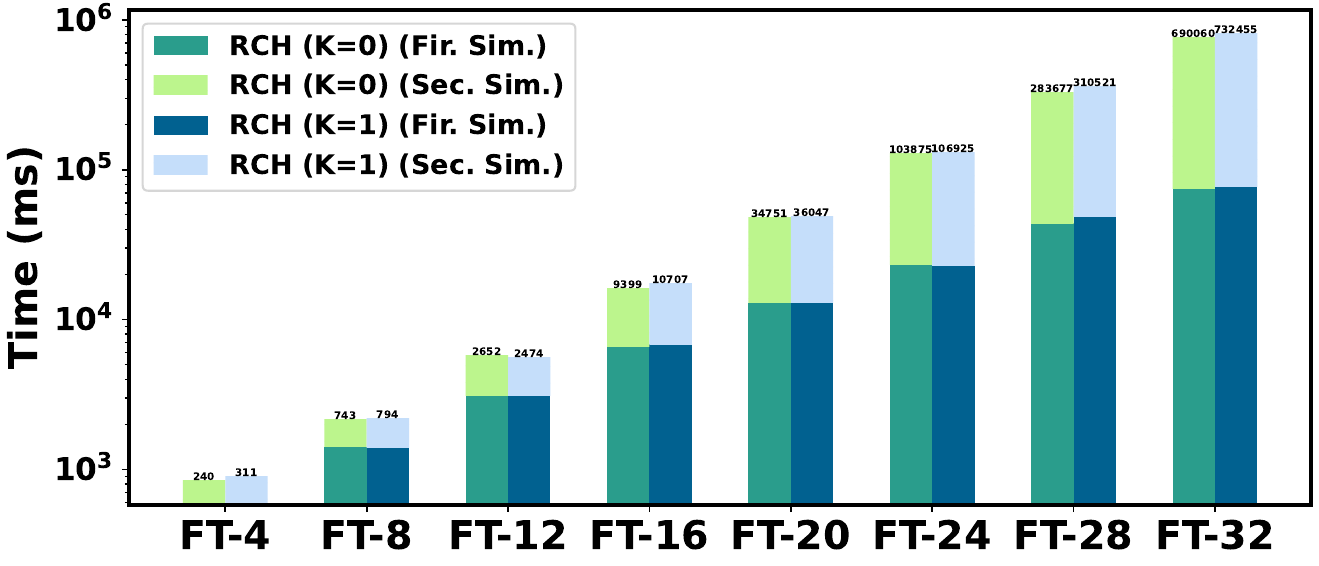}
    \vspace{-0.5em}
    \caption{Impact of net. scales on \system{} runtime in DCNs.}
    \label{fig:impact-of-scale}
    \vspace{-10pt}
\end{figure}

\para{Comparison.}
We evaluate CEL and CPR on real-world network configurations and find that CPR lacks support for  AS-path/community filters, the local preference modifier, and multi-protocol networks with the underlay–overlay architecture, while CEL similarly fails to handle regular-expression-based AS-path, community filters and the local-preference modifier present in these configurations. 

To verify accuracy, we use the example network in Figure~\ref{fig:example} and inject each real-world error from Table~\ref{tab:real-error} one at a time to evaluate the capabilities of existing tools. The results show that CPR can only diagnose and repair 5 out of the 10 injected errors, while CEL can diagnose 6. Notably, the errors these tools fail to handle are all commonly found in real DC-WAN and IPRAN configurations. In contrast, \system{} successfully supports all listed error types. Figure~\ref{fig:real-config-exp} shows the runtime of \system{} on real network configurations for various intent checks, including reachability, waypoint, and fault tolerance. The results demonstrate \system{}'s effectiveness and efficiency in handling feature-rich scenario. In the figure, the first simulation time refers to the runtime of the protocol simulator to generate data plane, common to all simulator-based tools, while the second simulation time corresponds to the selective symbolic simulation performed specifically by \system{}.



To further evaluate performance, we synthesize WAN configurations based on five real-world topologies from TopologyZoo, ranging from 33 to 155 nodes.
In each network, we randomly inject between one and five errors from the five error types supported by CEL and CPR. We define three sets of intents:
S1 (2 RCH + 2 WPT), S2 (6 RCH + 2 WPT), and S3 (10 RCH + 2 WPT).
Each injected error is crafted to violate at least one intent.
Details on the network scales, injected errors, and intents are summarized in Table~\ref{tab: result statistics table} in Appendix~\ref{appendix: exp}.

These synthesized configurations preserve realistic complexity, incorporating variations in network size, number of errors, and number of intents. Figure~\ref{fig:syn-config-diff-tools} compares the runtime of all tools across different network sizes and fault-tolerant scenarios. The results show that \system{} achieves over $>10$X speedup compared with CEL and CPR in diagnosing and repairing configuration errors under both no-link failure and single-link failure conditions. 
For large networks with 150+ nodes, CPR fails to complete the repair even for non-link-failure reachability, and CEL fails to complete the diagnosis of 1-link failure tolerance reachability.
\subsection{Scalability of \system{}}
\label{subsec: efficiency}

We use a collection of synthesized IPRAN (multi-protocol) and DCN (single-protocol) configurations with increasing size to test the scalability of our tool. 
We evaluate how the error type, error number, intent number, and network scale impact \system{}'s performance.




\para{The error type and error number have negligible impact on runtime.}
We choose IPRANs for evaluation due to their broader variety of injected errors compared with DCN.
We first vary the error type, while keeping the error number (\ie, 1) and the number of intents (\ie, 1) constant, to evaluate \system{}'s performance. 
We randomly inject one representative error from each IPRAN-related error category across 1K-, 2K-, and 3K-scale networks.
As Figure~\ref{fig:impact-of-err-category} shows, the diagnosis and repair time remain nearly constant across all error categories for each network. This is due to the fact that we model each contract as a Boolean condition, which requires nearly constant time to check the violations of different contracts. 

Since the error type has little impact on performance, we do not differentiate error type, and inject errors from all error types into the 1K-scale IPRAN with 10 intents. 
Figure~\ref{fig:impact-of-err-number} shows the average runtime when the error number varies. We can find that the runtime is nearly constant across different numbers of errors, indicating that the error number has negligible impact on runtime.

\para{The number of intents has a linear impact on runtime.}
We use a small-scale synthesized DCN (\ie, FT-8 with 80 nodes) for evaluation. 
Given that the error type and number are not fatal factors impacting performance, we randomly inject 10 errors into this network. 
We vary the number of intents that the network must satisfy. 
Figure~\ref{fig:impact-of-intents} shows the runtime required to diagnose and repair the network for both reachability and fault-tolerance reachability. We see a linear increase in repair times as the number of intents increases. This stems from two facts: (1) each new (unsatisfied) intent requires computing a new intent-compliant path, and (2) each intent-compliant path derives new contracts. Thus, the time required to compute the intent-compliant data plane and the time spent checking contract violations increase as the number of intents increases. 
Furthermore,  fault-tolerant reachability requires computing more paths and generating more contracts per intent, leading to a faster increase in runtime.

\para{The network scale has quadratic impact on runtime.}
We use synthesized DCNs for evaluation. 
We vary the number of nodes while keeping the number of intents constant (\ie, 10), and the results are shown in Figure~\ref{fig:impact-of-scale}. Although the overall runtime appears to grow exponentially with the network size, the majority of this increase comes from the first simulation, which is common to all simulation tools and should not be attributed to our system’s overhead.
In contrast, the second simulation, the core component of \system{}, exhibits quadratic growth. Results in Figure~\ref{fig:impact-of-err-category} (\ie, IPRAN) show a similar trend.
Additionally, we observe that the runtime of reachability and fault-tolerant reachability intents remains comparable. This is because computing one versus multiple intent-compliant paths introduces similar overhead in topologies that are highly symmetric and redundant, such as fat-trees. 

\vspace{-1em}
\section{Related Work} 
\vspace{-5pt}

\para{Network synthesis.}
Propane~\cite{beckett2016don} uses DFA to generate an intent-compliant data plane based on the intent. \system{} adopts this idea but proposes several principles to compute an intent-compliant data plane with small differences from the one generated by the erroneous configuration.
NetComplete~\cite{el2018netcomplete}, Merlin~\cite{soule2014merlin}, and AED~\cite{abhashkumar2020aed} symbolize parameters of the erroneous configuration and build a global constraint programming model to compute an intent-compliant configuration, which faces a search space explosion and potential unsatisfiable assignments due to contract conflicts.
To address these issues, \system{} uses contract-specific templates to enable per-violated-contract constraint programming.

\para{Network diagnosis and repair.}
Existing tools face various limitations.
CEL~\cite{gember2022localizing} uses an SMT-based model and computes the minimal correction set for error localization, but it cannot handle AS-path due to space explosion issues.
CPR~\cite{gember2017automatically} employs constraint programming to repair configuration errors but does not support local-preference, a commonly used attribute, due to graph abstraction.
ACR~\cite{acr-hotnets24} relies on coverage-based tools to identify erroneous configurations and repair errors using human experience, requiring multiple trials and potentially missing real errors. In contrast, \system{} proposes designs such as intent-compliant contract derivation, selective symbolic simulation, and contract-specific templates, enabling greater efficiency and accuracy.

\para{Configuration error localization.} 
Campion \cite{tang2021campion} localizes configuration errors by comparing the differences between the two given configurations, 
SelfStarter~\cite{kakarla2020finding} and Diffy~\cite{kakarla2024diffy} detect potential configuration errors by identifying outliers among a set of configurations. 
These works focus on finding errors across multiple router configurations, whereas our work targets automatically diagnosing and repairing intent violations within a single network configuration.

\para{Symbolic execution.}
A large body of prior work has applied symbolic execution to network verification or protocol verification.
For example, HSA~\cite{kazemian2012hsa}, APV~\cite{yang2015ap}, and SymNet~\cite{stoenescu2016symnet} symbolize the packet space to verify whether the network satisfies specific properties such as reachability or loop-free. 
Building on this idea, Hoyan~\cite{ye2020hoyan}, SRE~\cite{zhang2022sre}, and Expresso~\cite{wang2024expresso} symbolize the link-state space or external route space to verify whether the network satisfies specific properties under different environments. 
In contrast, our approach leverages symbolic execution to diagnose and repair configuration errors, shifting the focus from merely checking whether a property holds to identifying the root cause of violations and enabling corrective actions.

\vspace{-10pt}
\section{Limitations}
\vspace{-5pt}

\para{Inability to guarantee intent-compliant convergence in nondeterministic scenarios.} \system{} leverages the intent-compliant data plane to repair configuration errors. 
However, it cannot handle nondeterministic convergence scenarios, where a repaired configuration may have multiple converging states, such as BGP wedgies~\cite{bgp_wedgies}. In such cases, we cannot guarantee that the repaired configuration will always converge to the intent-compliant data plane.

\para{Non-minimal repair.} Our repair method does not guarantee minimal changes to the configuration. It simply minimizes differences in the data plane, assuming this leads to fewer violated contracts and smaller configuration updates. However, small data-plane changes may still trigger large configuration edits. Achieving minimal repair remains an open problem.


\para{Scalability for link-state protocols.} Our current solution for link-state protocols relies on global constraint solving and cannot leverage local templates, as used in path-vector protocols, to reduce the complexity and optimize performance.



\vspace{-10pt}
\section{Conclusion}
\vspace{-10pt}


This paper presents \system{}, a novel system for automatic routing configuration diagnosis and repair via the following new designs: 1) deriving intent-compliant data plane and contracts from erroneous configurations, 2) finding violated contracts and diagnosing errors via symbolic simulation, and 3) repairing errors with constraint programming.
Extensive experiments demonstrate its efficiency and efficacy. 

\textit{This work does not raise any ethical issues.}



\vspace{-1em}
\section*{Acknowledgments}
\vspace{-0.5em}
We sincerely thank our shepherd, Ryan Beckett, and the anonymous NSDI reviewers for their valuable feedback on this paper. 
We also thank Ruzica Piskac and Thomas Wies for their help during the preparation of this paper.
This work is supported in part by the National Key R\&D Program of China 2022YFB2901502, 
the National Natural Science Foundation
of China (Grant No. 62532010, 62572408, 62172345, and 62302409), and the Fundamental Research Funds for the Central Universities (Grant No. 20720250058 and 20720240049).




\ifcase \conftype
\bibliographystyle{ACM-Reference-Format}
\balance
\bibliography{ref}
\or

\bibliographystyle{plain}
\bibliography{ref}
\fi

\appendix
\newpage

\section{Outputs of Existing Tools}
\label{appendix}

\textbf{Batfish analyzed that intent 2 was violated, and returned a counter-example:}
\vspace{-1em}

\begin{figure}[H]
    \centering
    \includegraphics[width=\linewidth]{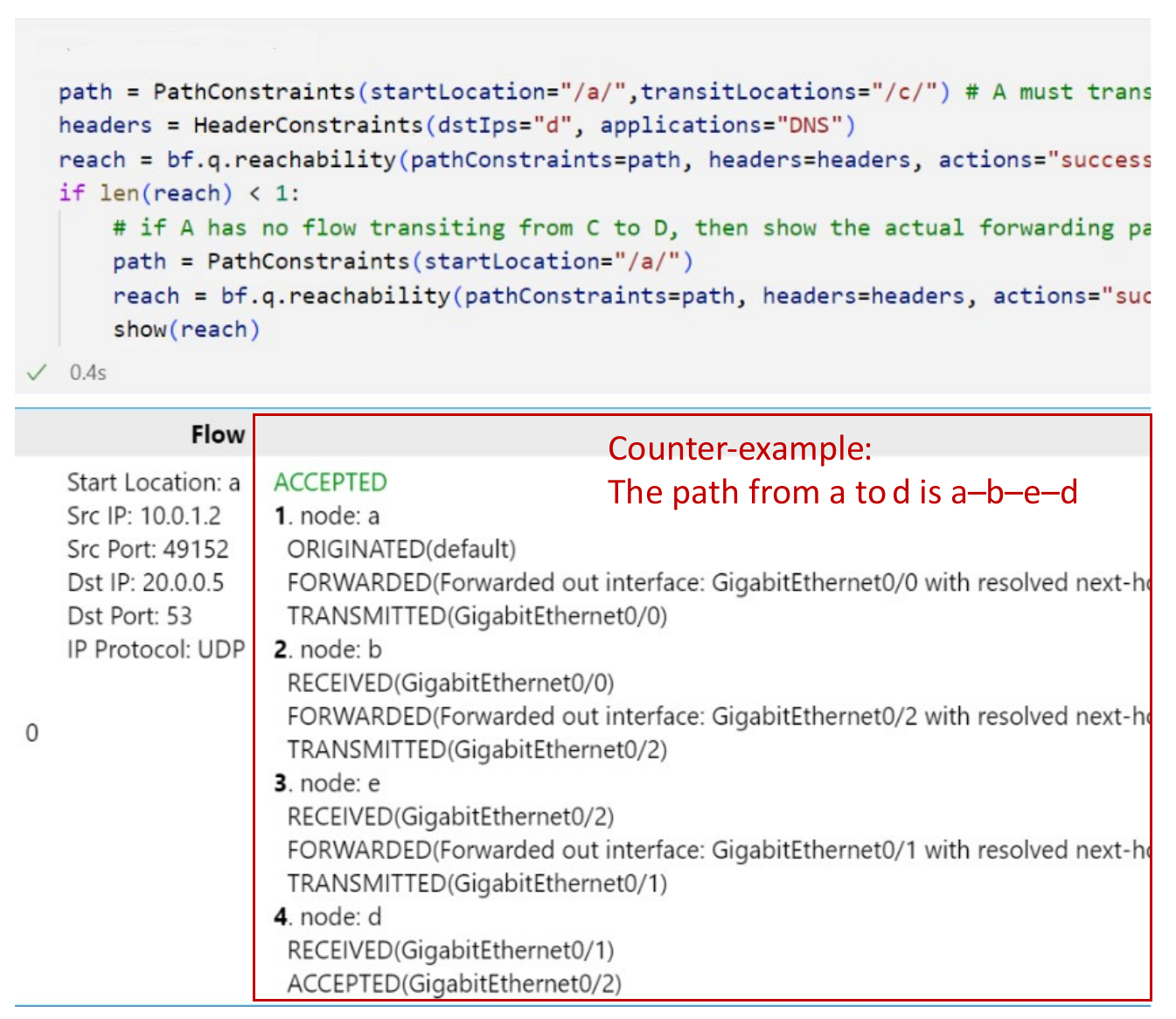}
    \caption{Batfish's output.}
    \label{fig:Batfish-output}
\end{figure} 
\vspace{-1em}

\para{Minesweeper verified that intent 2 was violated, and returned a counter-example:}
\vspace{-1em}

\begin{figure}[H]
    \centering
    \includegraphics[width=\linewidth]{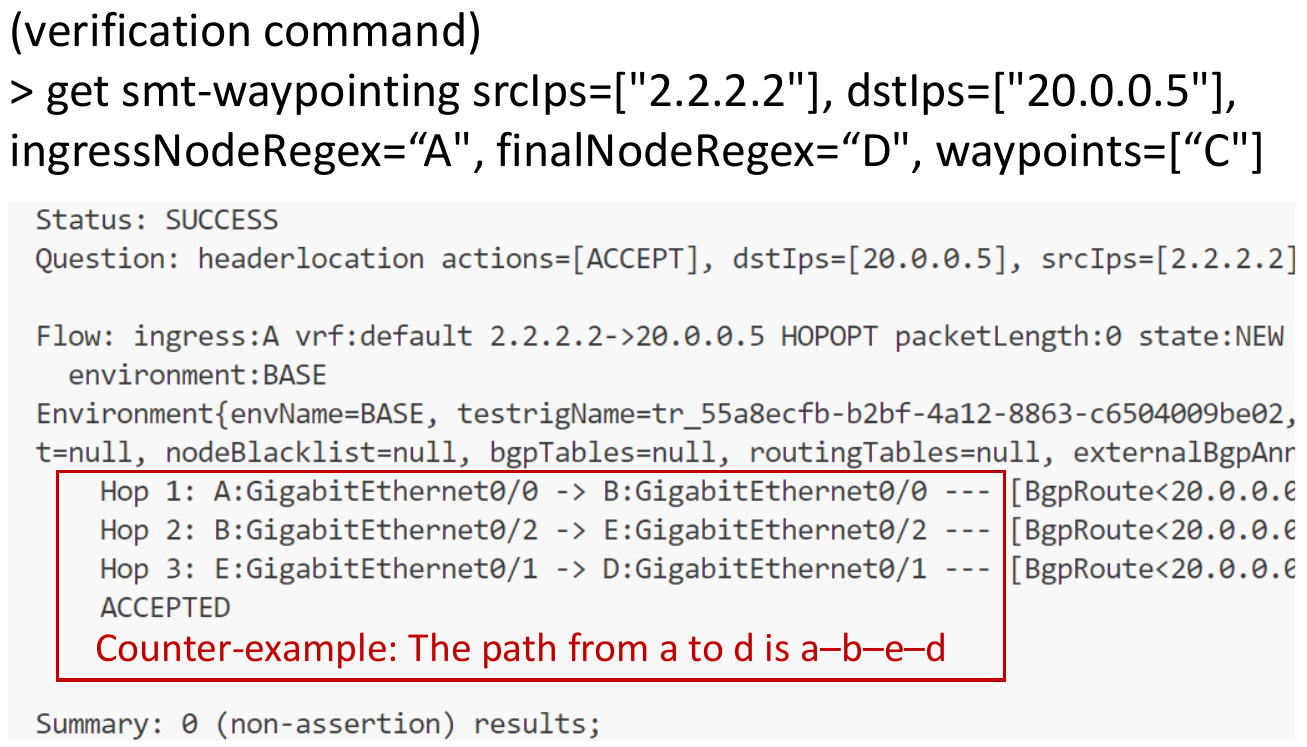}
    \caption{Minesweeper's output.}
    \label{fig:MS-output}
\end{figure} 
\vspace{-1em}

\para{CEL verified that intent 2 was violated, but only found one error out of two:}
\vspace{-1em}

\begin{figure}[H]
    \centering
    \includegraphics[width=\linewidth]{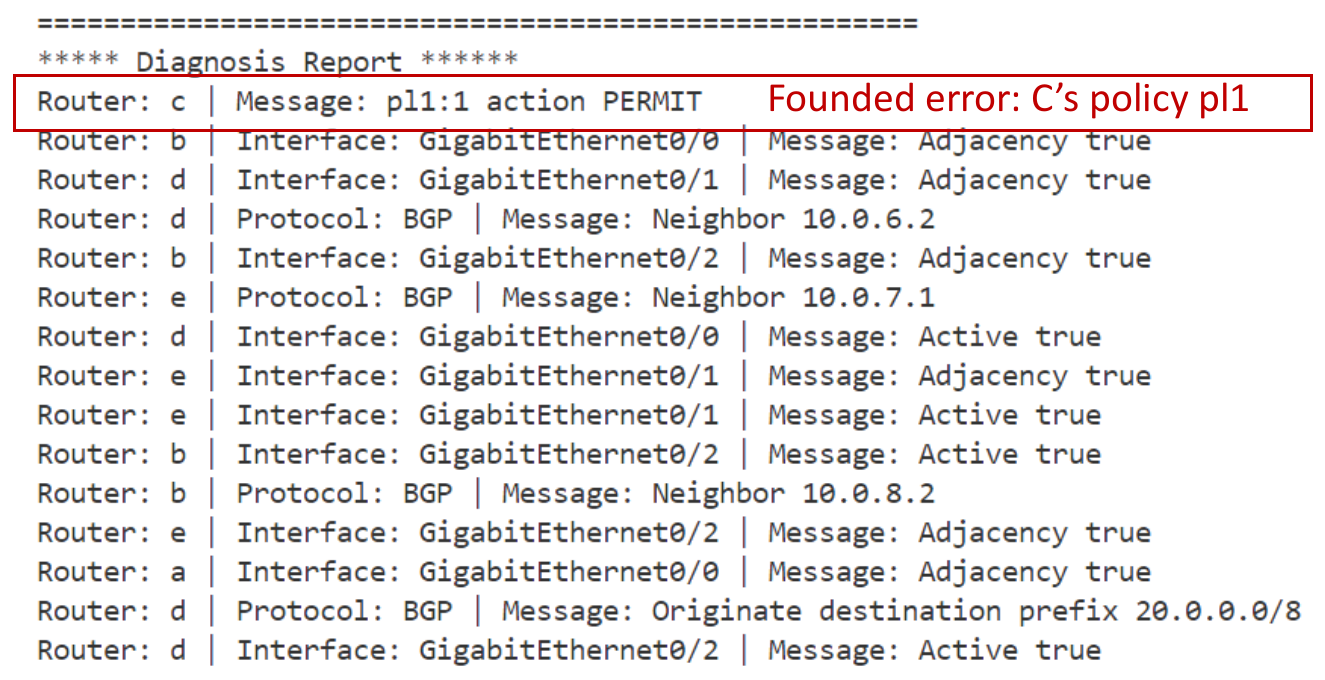}
    \caption{CEL's output.}
    \label{fig:CEL-output}
\end{figure} 

\para{CPR verified that intent 2 was violated, but found a patch that cannot fix any error:}
\vspace{-1em}

\begin{figure}[H]
    \centering
    \includegraphics[width=\linewidth]{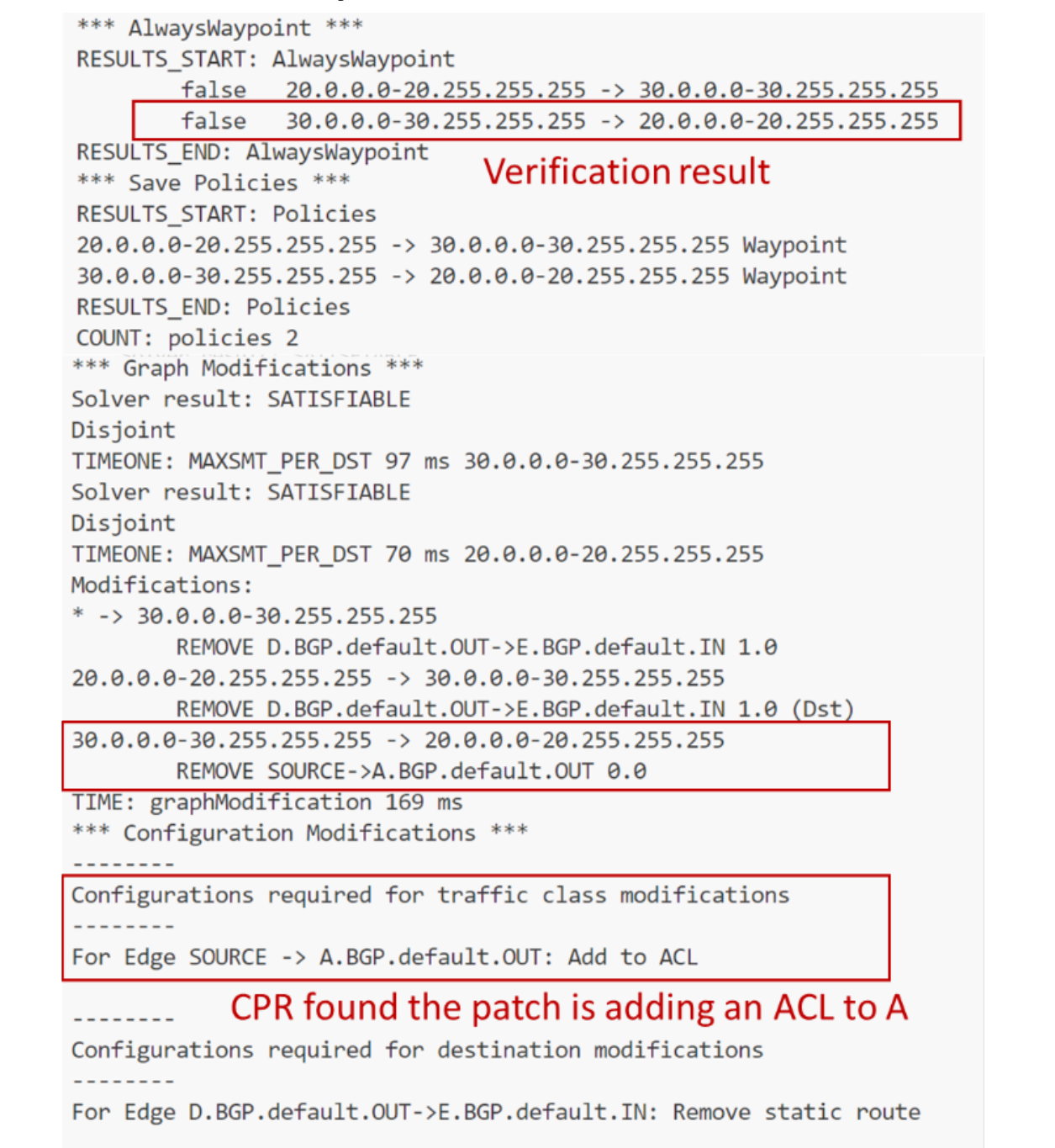}
        \caption{CPR's output.}
        \label{fig:CPR-output}
\end{figure}
\vspace{-1em}

\para{ACR depends on NetCov to identify suspicious lines first, but NetCov misses all erroneous for intent 2:}
\vspace{-1em}

\begin{figure}[H]
    \centering
    \includegraphics[width=\linewidth]{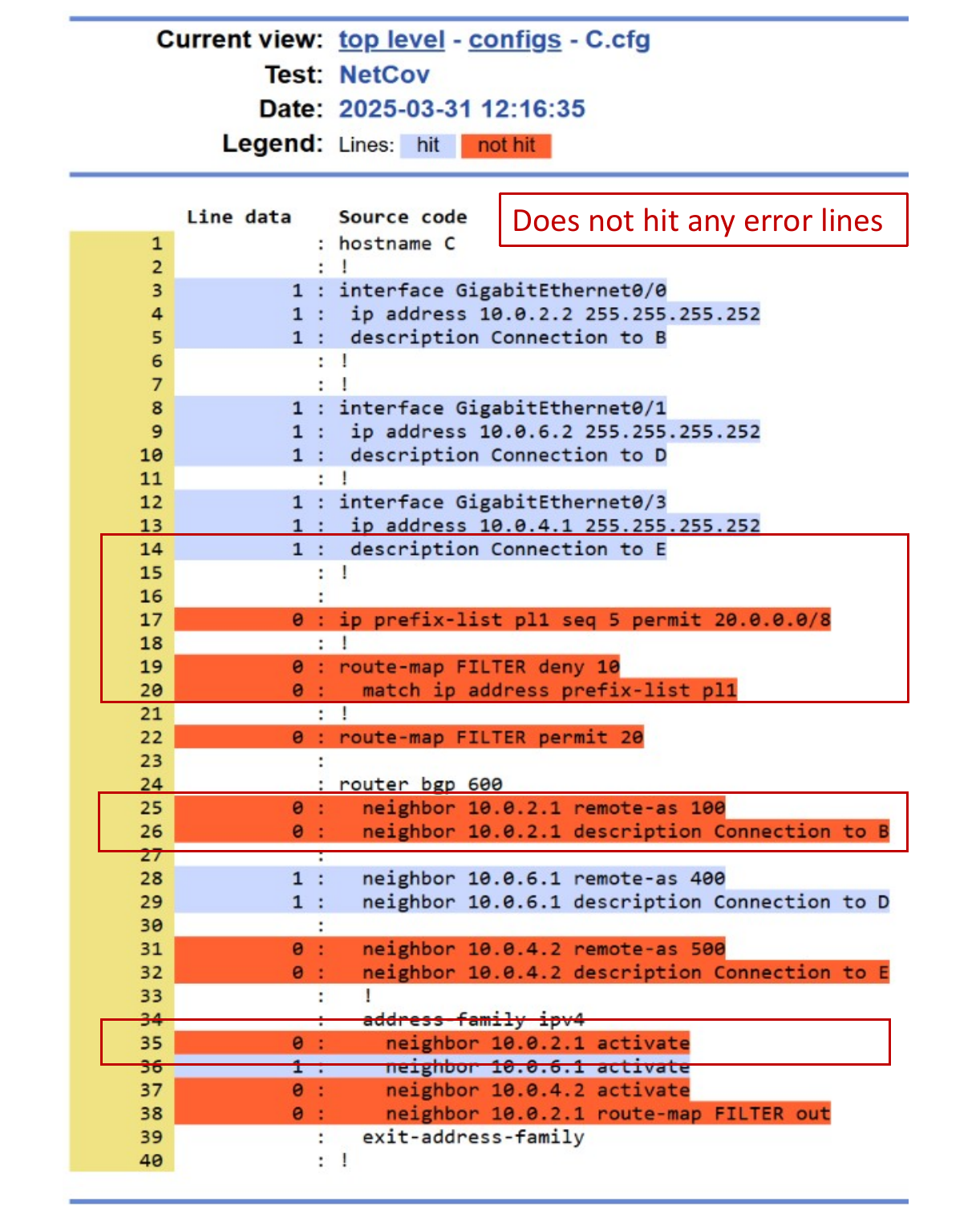}
        \caption{NetCov's output.}
        \label{fig:netcov-output}
\end{figure}


\section{Repair Templates of Contracts}
\label{appendix: templates}

We now present the repair templates for each contract in Table~\ref{tab: contracts}, illustrated using Cisco device configuration syntax.
For configurations from other vendors, manual translation or adaptation of the configuration may be necessary.
Note that the repair templates are not unique, and operators can tailor them according to their operational preferences.
For example, operators may prefer to use the peer group to organize neighbors, or modify attributes other than local-preference to adjust route selection and enforce desired priorities.

Each template consists of several configuration lines and contains parameter holes that must be completed according to the corresponding contract(s).
Specifically, holes marked with "[]" can be directly filled using contract parameters (\eg, the prefix of a route), whereas holes marked with "()" require resolution via constraint programming.

In the templates, lines marked with “+” indicate configurations that must be added or updated, whereas unmarked lines serve only as contextual anchors to indicate where the changes should be applied.

\para{isPeered contract.} 
To repair a \textit{isPeered(u,v)} contract, the peer configurations on both routers (u and v) must be updated.
The figure below shows the repair template on u, and the same applies to the other side.

The template consists of the minimal configuration necessary to establish a BGP peering session between two routers.
It specifies the remote AS number (ASN\textsubscript u), the update-source interface (IP\textsubscript u, usually the loopback interface), and the eBGP multihop parameter (HOP-CNT) when two routers establish a peering using non-adjacent interface IPs, followed by activation under the appropriate address family.

\begin{figure}[H]
    \centering
    \includegraphics[width=\linewidth]{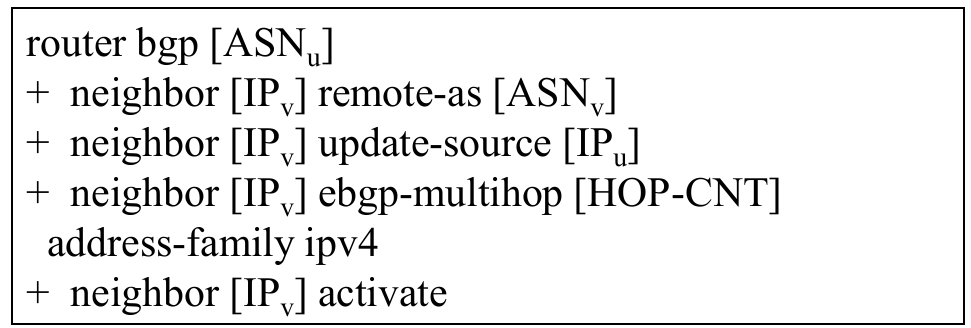}
\end{figure}

\para{isEnabled contract.} 
Repairing a \textit{isEnabled(u,v)} contract requires enabling the routing process on the corresponding adjacent interface for both sides. 
The figure below shows the repair template on router u.
The procedure differs depending on the protocol: for OSPF (left), it involves advertising the adjacent interface IP (IP\textsubscript u) to the corresponding OSPF process (OSPF-ID\textsubscript u), whereas for IS-IS (right), it requires enabling the IS-IS process (ISIS-ID\textsubscript u) on the adjacent interface (INF\textsubscript u).

\begin{figure}[H]
    \centering
    \includegraphics[width=\linewidth]{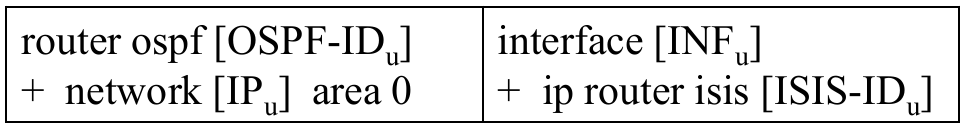}
\end{figure}

\para{isPreferred contract.}
To repair \textit{isPreferred(u, r, r')} contracts, we leverage the specific mechanisms of each routing protocol. In BGP, we utilize its fine-grained policy control to adjust the local preference of the non-preferred route $r'$, ensuring that $r$ is selected. 
The figure below shows the template on router u.

Specifically, we first find the policy (RM) that router u used to import routes from v (\ie, the sender of the non-preferred route $r'$). 
If such a policy does not exist in the original configuration, we create one.
Then, we employ a prefix list (PFL), a community list (CML), and an as-path list (APL) that precisely match the prefix, community, and AS-path of the non-preferred route $r'$.
Finally, we compute the concrete value of the match action (ACTION), sequence number (SEQ), and the local preference (LP), to ensure that $r'$ is less preferred than $r$.

\begin{figure}[H]
    \centering
    \includegraphics[width=\linewidth]{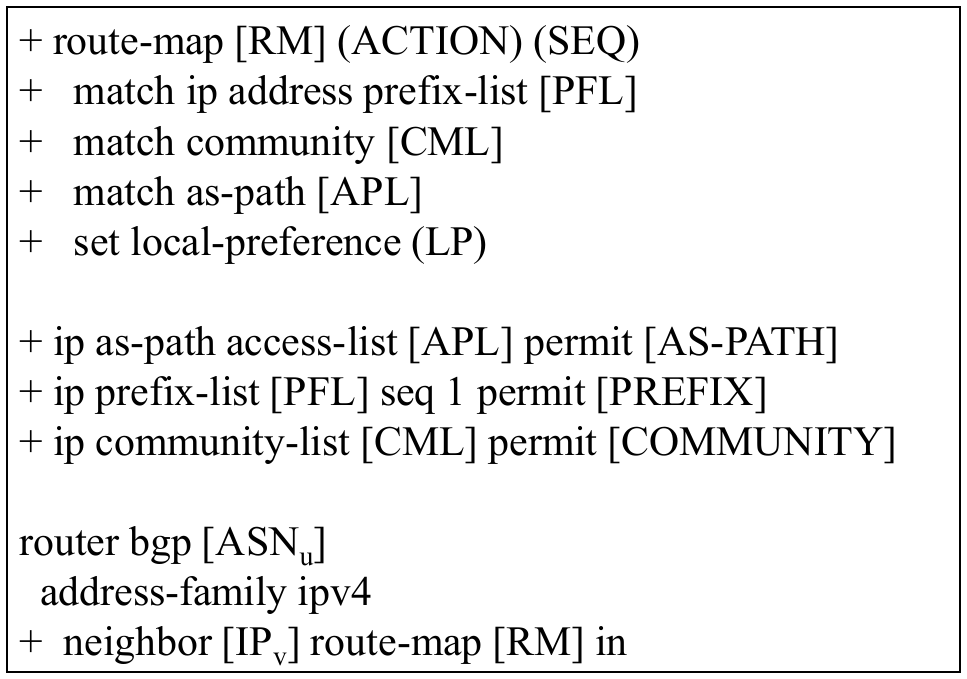}
\end{figure}

In link-state protocols such as OSPF and IS-IS, we repair this contract by recomputing the costs of all links to enforce the desired route preference using constraint programming. The figure below shows the template we used to adjust the link cost in a router.

\begin{figure}[H]
    \centering
    \includegraphics[width=\linewidth]{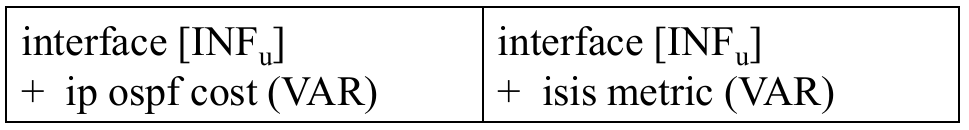}
\end{figure}


\begin{table*}[htbp]
\centering

\renewcommand{\arraystretch}{1.2} 
\begin{tabular}{|c|c|c|c|c|c|}
\hline
\textbf{Networks} & \textbf{Name} & \textbf{\#Node} & \textbf{\#Total Lines} & \textbf{Injected Error Type} & \textbf{\#Intents [RCH (K=0)/RCH (K=1)/WPT]} \\
\hline
\multirow{5}{*}{\parbox[c]{1.8cm}{\centering WAN\\(Topology Zoo)}}
 & Arnes     & 34   & 3.3K   & 1-1, 2-1, 2-3, 3-2 & 10 / 10 / 2 \\
\cline{2-6}
 & Bics      & 35   & 3.3K   & 1-1, 2-1, 2-3, 3-2 & 10 / 10 / 2 \\
\cline{2-6}
 & Columbus  & 70   & 6.3K   & 1-1, 2-1, 2-3, 3-2 & 10 / 10 / 2 \\
\cline{2-6}
 & Colt      & 155  & 13.4K  & 1-1, 2-1, 2-3, 3-2 & 10 / 10 / 2 \\
\cline{2-6}
 & GtsCe     & 149  & 13.3K  & 1-1, 2-1, 2-3, 3-2 & 10 / 10 / 2 \\
\hline
\multirow{3}{*}{IPRAN} 
 & IPRAN-1K  & 1006 & 176.1K & 1-1, 2-1, 3-1& 5 / - / - \\
\cline{2-6}
 & IPRAN-2K  & 2006 & 353.2K & 1-2, 2-1, 3-2     & 5 / - / - \\
\cline{2-6}
 & IPRAN-3K  & 3006 & 561.7K & 1-1, 2-3, 3-2     & 5 / - / - \\
\hline
\multirow{8}{*}{Fat-tree}
 & Fat-tree4  & 20   & 0.4K   & 1-1, 1-2         & 2 / 2 / - \\
\cline{2-6}
 & Fat-tree8  & 80   & 2.7K   & 1-1, 1-2         & 2 / 2 / - \\
\cline{2-6}
 & Fat-tree12 & 180  & 8.3K   & 1-1, 3-2         & 2 / 2 / - \\
\cline{2-6}
 & Fat-tree16 & 320  & 18.8K  & 1-1, 3-2    & 2 / 2 / - \\
\cline{2-6}
 & Fat-tree20 & 500  & 35.8K  & 1-1, 3-2         & 2 / 2 / - \\
\cline{2-6}
 & Fat-tree24 & 720  & 60.7K  & 1-2, 3-2         & 2 / 2 / - \\
\cline{2-6}
 & Fat-tree28 & 980  & 95.2K  & 1-2, 3-2         & 2 / 2 / - \\
\cline{2-6}
 & Fat-tree32 & 1280 & 140.8K & 1-2, 3-2         & 2 / 2 / - \\
\hline
\end{tabular}
\caption{Detailed statistics of synthetic configurations.}
\label{tab: result statistics table}
\end{table*}

\para{isEqPreferred contract.}
Repairing $isEqPreferred(u, r, r')$ contracts requires not only adjusting router configurations to assign equal preference to the routes (using the template of $isPreferred$ contracts), but also enabling the multipath policy on the corresponding routers.
The figure below shows the template that enables a router to utilize up to PATH-NUM paths for traffic forwarding, where the PATH-NUM is compliant with user intents.

\begin{figure}[H]
    \centering
    \includegraphics[width=\linewidth]{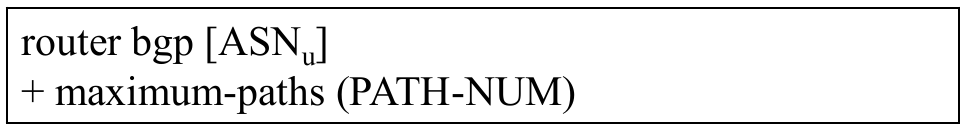}
\end{figure}

\para{isExport/isImport contract.}
Similar to repairing $isPreferred$ contracts in BGP, we leverage BGP's fine-grained policy control to export or import a route. The only difference is that, unlike in $isPreferred$, we do not modify the preference of the exported or imported route, as shown in the figure below.
The DIRECTION value (in or out) is determined by the contract type.

\begin{figure}[H]
    \centering
    \includegraphics[width=\linewidth]{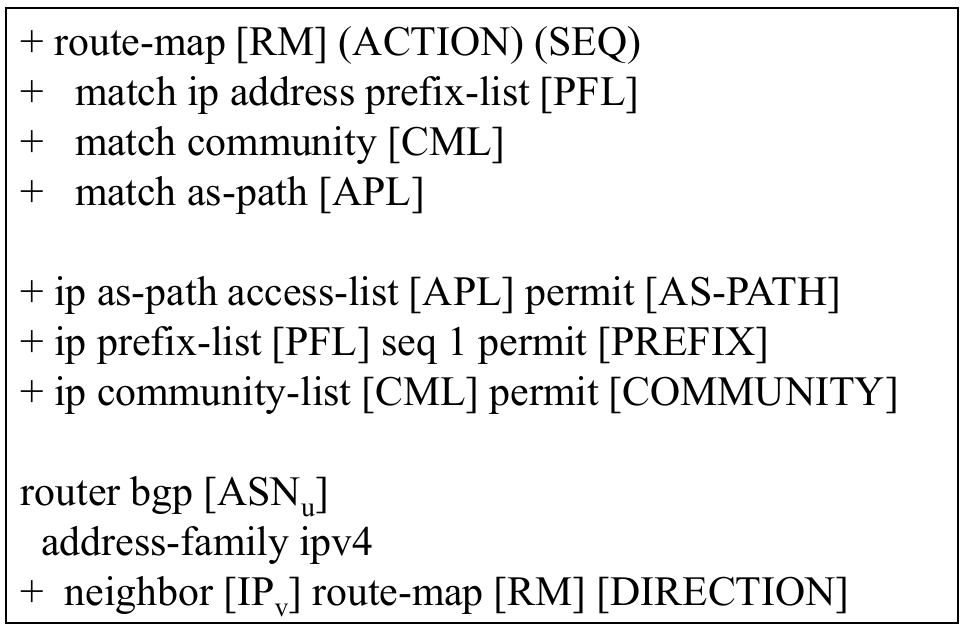}
\end{figure}

\para{isForwardedIn/isForwardedOut contract.}
Given an $isForwardedIn(u,p,v)$ contract, we repair it by adding/updating the access control list according to the packet $p$.
As shown in the figure below, we first locate the ACL on the adjacent interface through which router $u$ imports packets from $v$. If no such ACL exists in the original configuration, we create a new one.
Next, we insert a rule that exactly matches the prefix of packet $p$ before the existing rules of the ACL.
Finally, we determine the permit/deny action of the rule based on the value of the contract.

\begin{figure}[H]
    \centering
    \includegraphics[width=\linewidth]{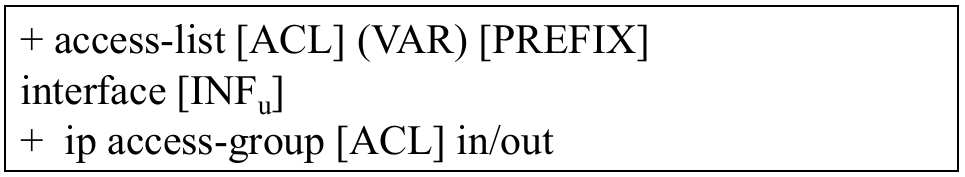}
\end{figure}


\section{Experimental Setup Details}
\label{appendix: exp}

Details of the synthesized networks for different network scales, injected errors, and intents are summarized in Table~\ref{tab: result statistics table}.

\end{document}